\documentclass[journal]{IEEEtran}

\usepackage{pifont}
\usepackage{epsfig,epsf,url,amssymb}
\usepackage{listings}  
\usepackage{subfigure}
\usepackage{amsmath}
\usepackage{rotating}
\usepackage{xspace}
\usepackage{paralist}
\usepackage{graphicx}

\usepackage{pifont}
\usepackage{epsfig,epsf,url,amssymb}
\usepackage[table]{xcolor}
\usepackage{multirow}
\usepackage{listings}  
\usepackage{subfigure}
\usepackage{amsmath}
\usepackage{rotating}
\usepackage{flushend}

\usepackage{color}
\newcommand{\delete}[1]{}

\usepackage{hhline}
\usepackage{textcomp,booktabs}

\newcommand{\generaldata}{\texttt{GOOGPLAY-FULL}\xspace}
\newcommand{\topkdata}{\texttt{GOOGPLAY-TOPK}\xspace}

\newcounter{ctProb}
{\begin{itemize}%
\setlength{\itemsep}{0pt}%
\setlength{\topsep}{0pt}%
\setlength{\partopsep}{0pt}%
\setlength{\parskip}{0pt}}%
{\end{itemize}}
{\begin{enumerate}%
\setlength{\itemsep}{0pt}%
\setlength{\topsep}{0pt}%
\setlength{\partopsep}{0pt}%
\setlength{\parskip}{0pt}}%
{\end{enumerate}}
\begin{document}

\title{A Longitudinal Study of Google Play}

\author{
   \IEEEauthorblockN{
   Rahul Potharaju\IEEEauthorrefmark{1},
   Mizanur Rahman\IEEEauthorrefmark{2},
   Bogdan Carbunar\IEEEauthorrefmark{2}}\\
   \IEEEauthorblockA{\IEEEauthorrefmark{1}Cloud and Information Services Lab, Microsoft}\\
   \IEEEauthorblockA{\IEEEauthorrefmark{2}Florida International University}
   \thanks{A preliminary version of this paper appears in ASONAM 2015.}
   \thanks{This research was supported in part by NSF grants 1527153, 1526494 and 1450619, and by DoD grant W911NF-13-1-0142.}
}

\maketitle

\begin{abstract}
The difficulty of large scale monitoring of app markets affects our
understanding of their dynamics.  This is particularly true for dimensions such
as app update frequency, control and pricing, the impact of developer actions
on app popularity, as well as coveted membership in top app lists. In this
paper we perform a detailed temporal analysis on two datasets we have collected
from the Google Play Store, one consisting of 160,000 apps and the other of
87,223 newly released apps. We have monitored and collected data about these
apps over more than 6 months. Our results show that a high number of these apps
have not been updated over the monitoring interval. Moreover, these apps are
controlled by a few developers that dominate the total number of app downloads.
We observe that infrequently updated apps significantly impact the median app
price. However, a changing app price does not correlate with the download
count. Furthermore, we show that apps that attain higher ranks have better
stability in top app lists. We show that app market analytics can help detect
emerging threat vectors, and identify search rank fraud and even malware.
Further, we discuss the research implications of app market analytics on
improving developer and user experiences.
\end{abstract}


\section{Introduction}\label{sec:introduction}

The revolution in mobile device technology and the emergence of ``app
markets'', have empowered regular users to evolve from technology consumers to
enablers of novel mobile experiences.  App markets such as Google Play provide
new mechanisms for software distribution, collecting software written by
developers and making it available to smartphone users. This centralized
approach to software distribution contrasts the desktop paradigm, where users
obtain their software directly from developers.

Developers and users play key roles in determining the impact that market
interactions have on future technology. However, the lack of a clear
understanding of the inner workings and dynamics of popular app markets,
impacts both developers and users. For instance, app markets provide no
information on the impact that developer actions will likely have on the
success of their apps, or guidance to users when choosing apps, e.g., among
apps claiming similar functionality.

This situation is exploited however by fraudulent and malicious developers.
The success of Google Play and the incentive model it offers to
popular apps~\footnote{ Google offers financial incentives for contribution to
app development, by making revenue sharing transparent for developers (70-to-30
cut, where developers get 70\% of the revenue).}, make it an appealing target
for fraudulent and malicious behaviors. Fraudulent developers have been shown
to attempt to engineer the search rank of their apps~\cite{GPlay.Fake}, while
malicious developers have been shown to use app markets as a launch pad for
their
malware~\cite{Malware.PCWorld,Malware.PCMag,Malware.Fortune,Malware.Forbes}.

\noindent{\bf Contributions}.
In this article we seek to shed light on the dynamics of Google Play, the most
popular Android app market. We report results from one of the first
characteristic studies on Google Play, using real-world time series data.  To
this end, we have developed iMarket, a prototype app market crawling system. We
have used iMarket to collect data from more than 470,000 Google Play apps, and
daily monitor more than 160,000 apps, over more than 6 months.

We use this data to study two key aspects of Google Play.  First, we seek to
understand the dynamics of the market in general, from an application and
developer perspective. For this, we evaluate the frequency and characteristics
of app updates (e.g., their effects on bandwidth consumption), and use the
results to determine if developers price their apps appropriately. We show that
only 24\% of the 160,000 app that we monitored have received an update within 6
months, and at most 50\% of the apps in any category have received an update
within a year from our observation period.  We conclude that market inactivity
has a significant impact on the price distribution. Therefore, while pricing is
an important and complex task, relying on statistics computed on the entire
population (as opposed to only active apps) may mislead developers, e.g., to
undersell their apps ($\S$\ref{sec:staleness}). Also, we show that typical app
update cycles are bi-weekly or monthly. More frequently updated apps (under
beta-testing or unstable) can impose substantial bandwidth overhead and expose
themselves to negative reviews ($\S$\ref{sec:updates}).

To evaluate the developer impact, we first seek to verify our hypothesis that a
few developers control the app market supply. Our analysis reveals however that
developers that create many applications are not creating many popular
applications. Instead, we discovered that a few elite developers are
responsible for applications that dominate the total number of downloads
($\S$\ref{sec:developer:impact}). Second, we evaluate the impact of developer
actions on the popularity of their apps. We show that few apps frequently
change prices, and with every subsequent software update, a developer is more
likely to decrease the price. However, changing the price does not show an
observable association with the app's download
count($\S$\ref{sec:developer:impact}).

A second key aspect of Google Play that we study is the temporal evolution of
top-k ranked lists maintained by the market. Top-k lists reveal the most
popular applications in various categories. We show that a majority of apps in
top-k app lists follow a ``birth-growth-decline-death'' process: they enter and
exit from the bottom part of a list.  Apps that attain higher ranks have better
stability in top-k lists than apps that are at lower ranks
($\S$\ref{sec:topk}).

\noindent
{\bf Impact of the study}.
A longitudinal study of Google Play app metadata can provide unique information
that is not available through the standard approach of capturing a single app
snapshot. Features extracted from a longitudinal app analysis (e.g.,
permission, price, update, download count changes) can provide insights into
fraudulent app promotion and malware indicator behaviors. For instance, spikes
in the number of positive or negative reviews and the number of downloads
received by an app can indicate app search optimization campaigns launched by
fraudsters recruited through crowdsourcing sites. Frequent, substantial app
updates may indicate Denial of Service (DoS) attacks, while permission changes
can indicate benign apps turning malicious see
$\S$~\ref{sec:discussion:threat}. Features extracted from a longitudinal app
monitoring can be used to train supervised learning algorithms to detect such
behaviors.

In addition, a detailed longitudinal study of Google Play apps can improve
developer and user experiences. For instance, app development tools can help
developers optimize the success of their apps. Such tools can integrate
predictions of the impact that price, permissions and code changes will have on
the app's popularity, as well as insights extracted from user reviews. In
addition, visualizations of conclusions, and analytics similar to the ones we
perform in this paper, can help users choose among apps with similar claimed
functionality.

We include a detailed discussion of the applicability and future research
directions in app market analytics in $\S$\ref{sec:discussion}.

\section{Related Work}
\label{sec:related}

This article extends our preliminary work~\cite{CP15} with iMarket, the market
crawler we developed and used to collect the data, new scores to evaluate the
evolution and variability of top-k lists and new experiments and evaluations.

Viennot et al.~\cite{VGN14} developed PlayDrone, a crawler to collect Google
Play data. Their main finding is that Google Play developers often include
secret key information in the released apps, making them vulnerable to attacks.
They further analyze the data and show that Google Play content evolves quickly
in time, that 25\% of apps are clones, and that native experience correlates
strongly to popularity. The analysis is performed over data collected for 3
non-contiguous months (May/June 2013 and November 2013). In contrast, our
analysis is performed over apps monitored daily over more than 6 months.
Furthermore, our analysis includes orthogonal app market dynamics dimensions,
that include the frequency and cycles of app updates, the developer impact and
control on the app market, and the dynamics of top-k lists.

Zhong and Michahelles~\cite{ZM13} analyze a dataset of Google Play
transactions, and suggest that Google Play is more of a ``Superstar'' market
(i.e., dominated by popular hit products) than a ``Long-tail'' market (i.e.,
where unpopular niche products contribute to a substantial portion of
popularity).  In addition, Zhong and Michahelles~\cite{ZM13} show that certain
expensive professional apps attract disproportionately large sales.  This is
consistent with our finding that a few developers are responsible for the most
popular apps.

M{\"o}ller et al.~\cite{MMDRK12} use an app they posted on Google Play to study
the correlation between published updates and their actual installations. They
show that 7 days after a security update is published, almost half of the
app's users still use an older, vulnerable version.
Liu et al.~\cite{LAC12} use a dataset of 1,597 ranked mobile apps to conclude
that the ``freemium'' strategy is positively associated with increased sales
volume and revenue of the paid apps.  Moreover, they show that free apps that
rate higher contribute to higher revenue for the paid version. We note that our
work studies a multitude of previously unanswered questions about Google Play,
regarding app update frequency and pricing appropriateness, and the evolution
of top-k lists.

Petsas et al.~\cite{PPPMK13} explored mobile app markets in the context of 4
providers, that do not include Google Play. They show that the distribution of
app popularity deviates from Zipf, due in part to a strong temporal affinity of
user downloads to app categories. They show  that on the markets they studied,
paid apps follow a different popularity distribution than free apps. In
contrast, our work exclusively analyzes Google Play, the most popular Android
app market. In addition, we focus on different dimensions: (i) app update
frequency and its effect on app pricing and resource consumption, (ii) the
control of the market and the effect of developer actions on the popularity of
their apps and (iii) the evolution in time of top apps and top-k app lists.

Xu et al.~\cite{XEGMPV11} use IP-level traces from a tier-1 cellular network
provider to understand the behavior of mobile apps.  They provide an orthogonal
analysis of spatial and temporal locality, geographic coverage, and daily usage
patterns.

Security has been a theme in the large scale collection of mobile apps.
Previous work includes malware detection~\cite{ZWZJ12}, malware
analysis~\cite{ZJ12}, malicious ad libraries~\cite{GZJS12}, vulnerability
assessment~\cite{EOMC11}, overprivilege identiﬁcation~\cite{FCHSW11} and
detection of privacy leaks~\cite{EGCCJMS10}. While in this paper we focus on
the different problem of understanding the dynamics of Google Play, we also
introduce novel mobile app attacks.


\section{Google Play Overview}

\noindent
\textbf{App Distribution Channel}: Google Play is the app distribution channel
hosted by Google. Each app submitted by a developer gets an entry on the market
in the form of a webpage, accessible to users through either the Google Play
homepage or the search interface. This webpage contains meta-information that
keeps track of information pertaining to the application (e.g., name, category,
version, size, prices). In addition, Google Play lists apps according to
several categories, ranging from ``Arcarde \& Action'' to ``Weather''. Users
download and install apps of interest, which they can then review. A review has
a rating ranging from 1 to 5. Each app has an \textit{aggregate rating}, an
average over all the user ratings received. The app's webpage also includes its
usage statistics (e.g., rating, number of installs, user reviews). This
information is used by users when they are deciding to install a new
application.

\noindent
\textbf{App Development}: In order to submit apps to Google Play, an Android
developer first needs to obtain a publisher account for a one-time fee of \$25.
The fee encourages higher quality products and reduces spam~\cite{googledev}.
Google does not limit the number of apps that can be submitted by developers.
As a measure to reduce spam, Google recently started the
Bouncer~\cite{googlebouncer} service, which provides automated scanning of
applications on Google Play for potential malware.  Developers can sell their
apps for a price of their choice, or distribute them for free.

\noindent
\textbf{Permission Model}: Android follows the
Capability-based~\cite{capsbasedsystems} security model. Each app must declare
the list of capabilities (permissions) it requires in a manifest file called
\textit{Android-Manifest.xml}. When a user downloads an app through the Google
Play website, the user is shown a screen that displays the permissions
requested by the application. Installing the application means granting the
application all the requested permissions i.e. an \textit{all-or-none}
approach.

\section{Data Collection}
\label{sec:data}

We use \textit{snapshot} to refer to the entire state of the market i.e., it
contains meta information of all apps. We first describe iMarket, our app market
crawler, then describe the datasets that we collected from Google Play.

\subsection{The iMarket Crawler}

\begin{figure}
\centering
{\includegraphics[width=0.49\textwidth]{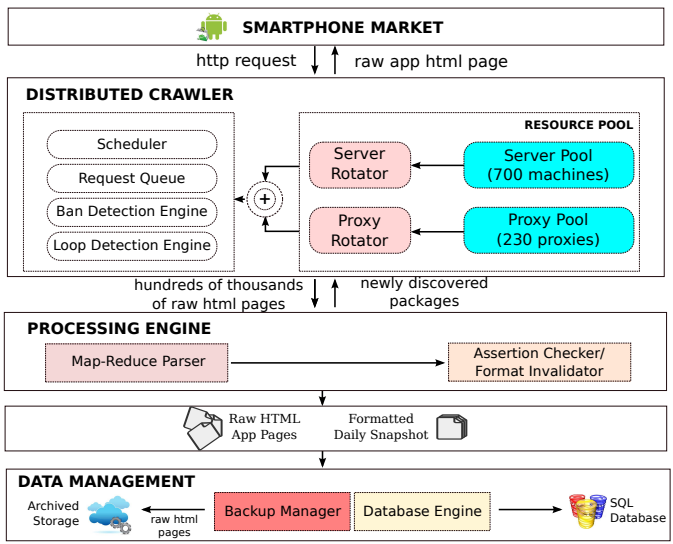}}
\caption{Architecture of iMarket, the developed GooglePlay
crawler. It consists of a distributed crawler, processing engine and data
management components.}
\label{fig:imarket}
\vspace{-15pt}
\end{figure}

\begin{figure*}
\centering
\subfigure[]
{\label{fig:freepaid}{\includegraphics[width=0.49\textwidth]{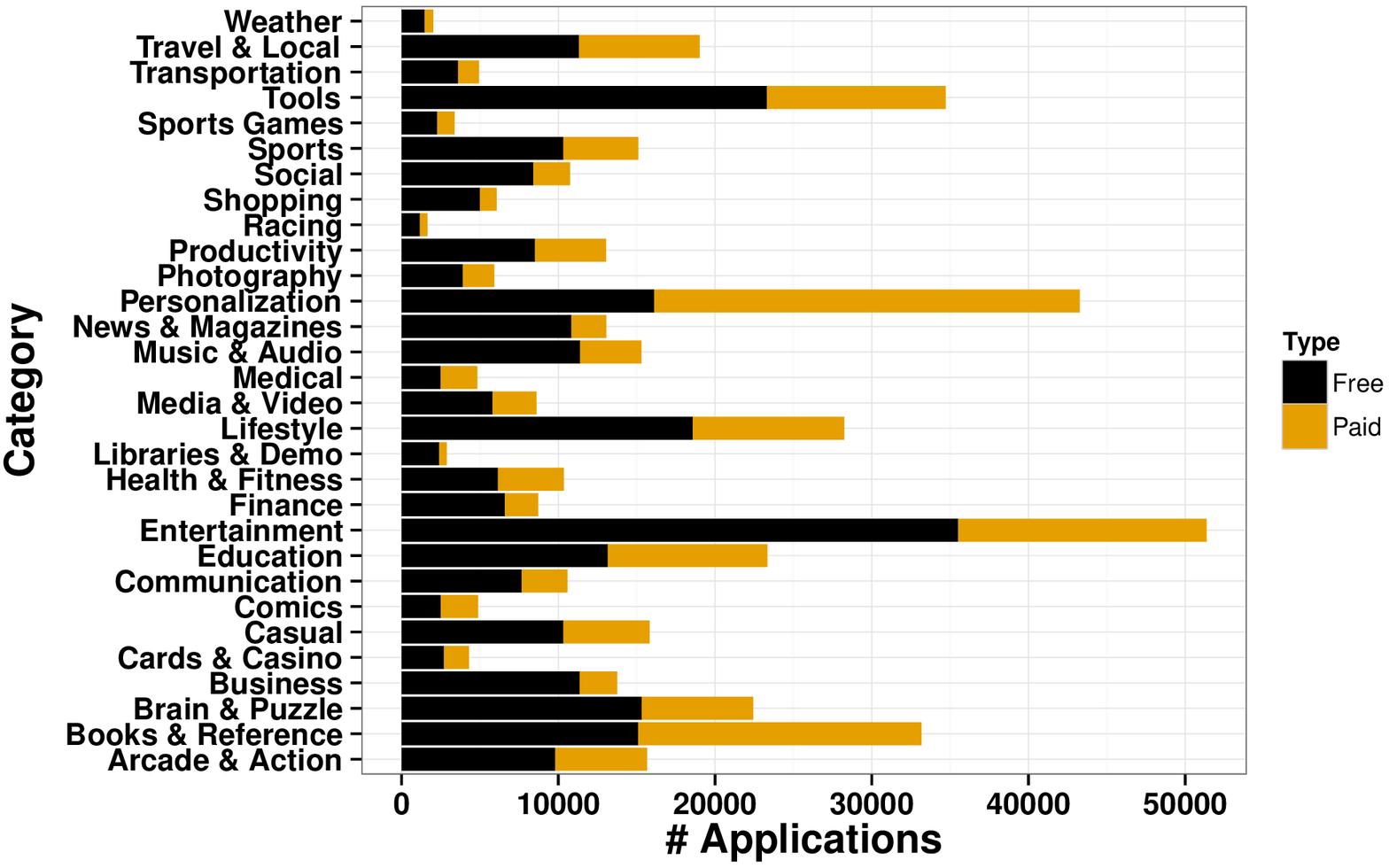}}}
\subfigure[]
{\label{fig:freepaid:fresh}{\includegraphics[width=0.41\textwidth]{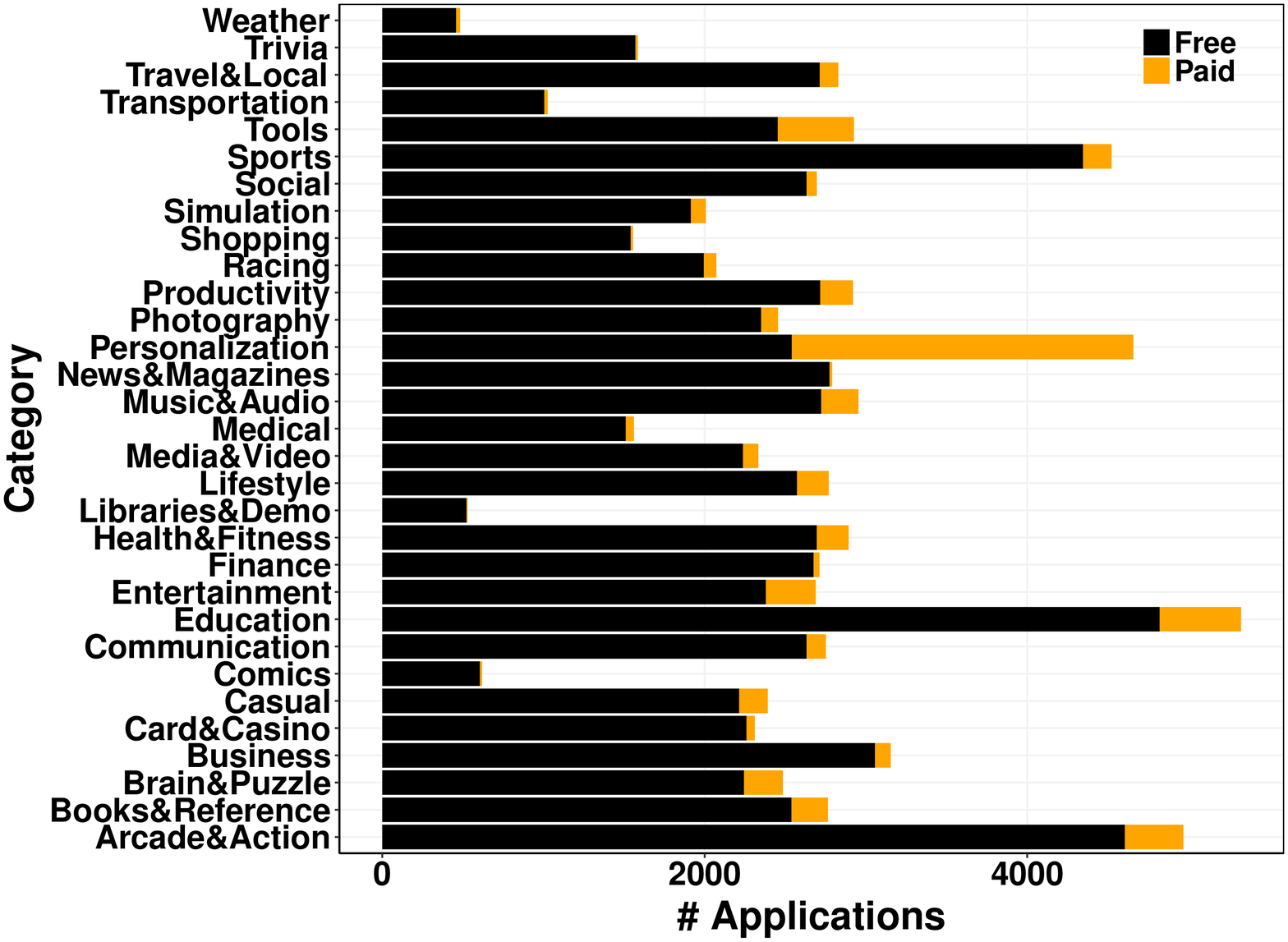}}}
\caption{Distribution of free vs. paid apps, by category, for (a) dataset.2012
and dataset.14-15. The number of free apps exceeds the number
of paid ones especially in dataset.14-15. We conjecture that this occurs due to
user tendency to install more free apps than paid apps. Since 2012, developers
may have switched from a direct payment model for paid apps, to an ad based
revenue model for free apps.}
\vspace{-5pt}
\end{figure*}

iMarket, our prototype market crawling system (see Figure~\ref{fig:imarket} for
an overview) consists of three main components. First, the \textit{Distributed
Crawler} component, which is responsible for crawling the target market and
collecting information on various apps that are accessible from the current
geographical location.  We initially leveraged hundreds of foreign proxies to
address challenge 3 above.  However, we later decided to rely only on local
US-based proxies for stability reasons. While this trades-off completeness for
consistency, having continuous information about a few apps improves
the accuracy of most statistical inference tasks compared to having discrete
information about hundreds of thousands of apps.

To seed our distributed crawler, we initially ran it using a list consisting of
about 200 randomly hand-picked apps from different categories. To address
Challenge 1, our app discovery process is designed as follows: After retrieving
each page, the ``Similar Apps'' portion of the raw HTML page is parsed to
obtain a new list of packages. These packages are queued for crawling and
simultaneously appended to the previous day's package list.  We have also
detected a ban detection engine in place that deactivates servers once it
observes a threshold number of ``404 Not Found'' messages (Challenge 2) from
the market provider.

The second component, the ``Processing Engine'' contains a \textit{Map-Reduce
Parser} component that uses the map-reduce paradigm~\cite{dean2008mapreduce} to
handle parsing of hundreds of thousands of raw HTML app pages. In the ``map''
stage, a chunk of files ($\approx$10K) are mapped onto each of the 700 machines
and a parser (written in Python) parses these HTML and extracts the meta
information.  In the ``reduce'' stage, these individual files are combined into
a single file and de-duplicated to maintain data integrity. This stage takes
$\approx$1-1.5 hours. After constructing the aggregate file, we address
Challenge 3 using the assertion checker that takes a \textit{best-effort}
approach to ensure that all the information has been correctly parsed from the
raw files. Note that despite our best-effort approach, our dataset still
contained some missing information due to temporary unavailability/maintenance
of servers.


The third, ``Data Management" component, archives the raw HTML pages
($\approx$14 GB compressed/day) in a cloud storage to support any ad hoc
processing for other tasks (e.g., analyzing HTML source code complexity) and
subsequently removed from the main servers. To address Challenge 4, the
formatted daily snapshot ($\approx$200 MB/day) is then inserted into a database
to support data analytics. We setup the relevant SQL Jobs to ensure that
indexes are re-built every two days --- this step significantly speeds up SQL
queries. Our six months of archived raw files consume $\approx$7 TB of storage
and the database consumes $\approx$400 GB including index files.

\section{Data}
\label{sec:data:sets}

\begin{figure*}
\centering
\subfigure[]
{\label{fig:agesince:2012}{\includegraphics[width=0.405\textwidth,height=2.35in]{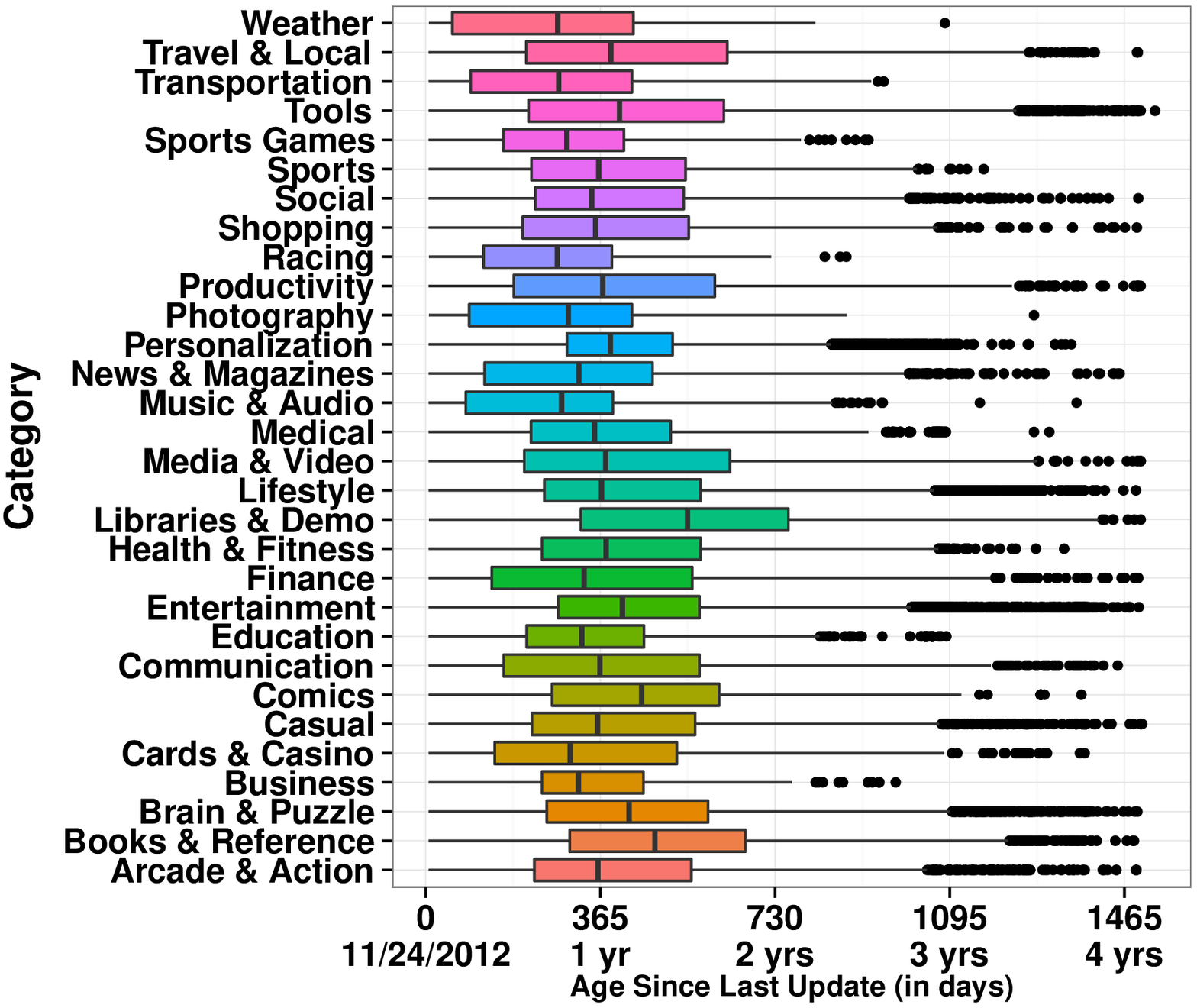}}}
\subfigure[]
{\label{fig:agesince:14-15}{\includegraphics[width=0.37\textwidth,height=2.29in]{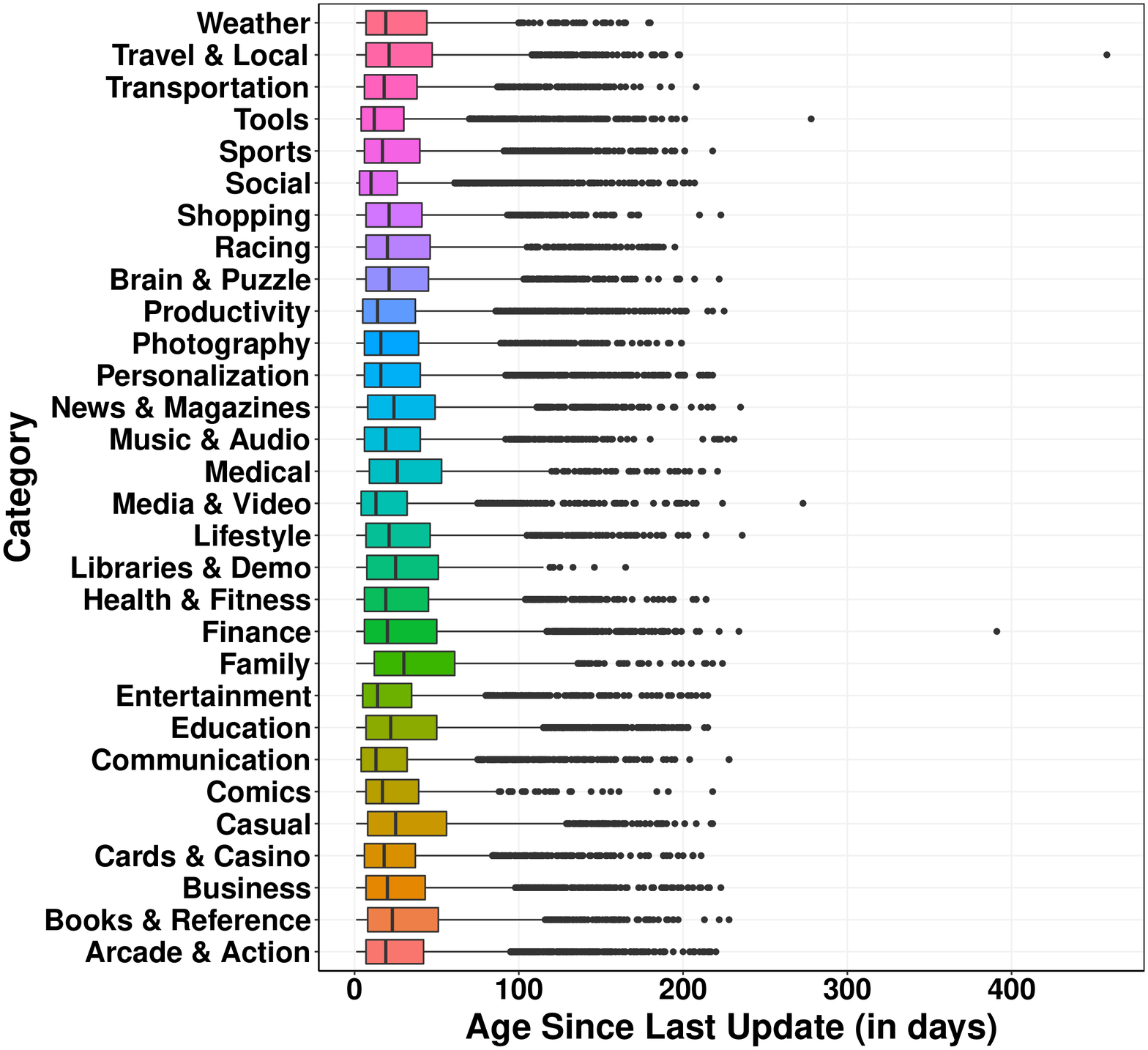}}}
\caption{Box and whiskers plot of the time distribution from the last update,
by app category, for (a) dataset.2012: at most 50\% of the apps in each
category have received an update within a year and (b)
dataset.14-15: at most 50\% of the apps in each category have received an
update within 35 days. {\bf This may occur since new apps are likely to have
more bugs and receive more attention from developers.}}
\label{fig:agesince}
\vspace{-15pt}
\end{figure*}

We used iMarket to collect two Google Play datasets, which we call 
dataset.2012 and dataset.14-15.

\subsection{Dataset.2012}

We have used a total of 700 machines~\footnote{We have used 700
machines, each with a different IP address and from a different subnet, in
order to avoid getting banned during the crawling process.} for a period of
7.5 months (February 2012 - November 2012) to collect data from 470,000 apps.
The first 1.5 months are the ``warm up'' interval. We do not consider data
collected during this period for subsequent analysis. Instead, we focus on a
subset of 160K apps for which we have collected the following data:

\noindent
\textbf{\generaldata}: We used \textit{iMarket} to take
daily snapshots of Google Play store from April - November, 2012.  For each
app, we have daily snapshots of application meta-information consisting of the
developer name, category, downloads (as a range i.e., 10-100, 1K-5K etc.),
ratings (on a 0-5 scale), ratings count (absolute number of user ratings), last
updated timestamp, software version, OS supported, file size, price, url and
the set of permissions that the app requests. Figure~\ref{fig:freepaid} shows the
distribution of apps by category. While overall, the number of free apps exceed
the number of paid apps, several popular categories such as ``Personalization''
and ``Books \& References'' are dominated by paid apps.

\noindent
\textbf{\topkdata}: Google publishes several lists, e.g., \textit{Free}
(most popular apps), \textit{Paid} (most popular paid), \textit{New (Free)}
(newly released free apps), \textit{New (Paid)} (newly released paid) and \textit{Gross}
(highly grossing apps). Each list is divided into
$\approx$20 pages, each page consisting of 24 apps. These lists are
typically updated based on application arrival and the schedule of Google's
ranking algorithms. Since we cannot be notified when the list
changes, we took hourly snapshots of the lists. Our \topkdata
consists of hourly snapshots of five top-k lists ($\approx$ 3000 apps) from
Jul-Nov, 2012 ($\approx$2880 hours worth of data).

\subsection{Dataset.14-15}

Further, we have used a dataset of more than 87,000 newly released apps that we
have monitored over more than 6 months~\cite{RRCC16}. Specifically, we have
collected newly released apps once a week, from Google Play's ``New Release''
links, to both free and paid apps.  We have validated each app based on the
date of the app's first review: we have discarded apps whose first review was
more than 40 days ago. We have collected 87,223 new releases between July and
October 2014, all having less than 100 reviews.

We have then monitored and collected data from these 87,223 apps between
October 24, 2014 and May 5, 2015. Specifically, for each app we captured
``snapshots'' of its Google Play metadata, twice a week. An app snapshot
consists of values for all its time varying variables, e.g., the reviews, the
rating and install counts, and the set of requested permissions. For each of
the $2,850,705$ reviews we have collected from the $87,223$ apps, we recorded
the reviewer's name and id, date of review, review title, text, and rating.

Figure~\ref{fig:freepaid:fresh} shows the distribution of apps by category.
With the exception of the ``Personalization'' category, the number of free apps
significantly exceeds the number of paid apps. We have observed that
consistently through our collection effort, we identified fewer top paid than
free new releases. One reason for this may be that users tend to install more
free apps than paid apps. Thus, not only developers may develop fewer paid
apps, but paid apps may find it hard to compete against free versions. We
note that free apps bring revenue through ads.


\section{Popularity and Staleness}
\label{sec:pop:staleness}

We first evaluate the fraction of apps that are active, and
discuss the implications this can have on app pricing. We then classify apps
based on their popularity, and study the distribution of per-app rating counts.
Finally, we study the frequency of app updates for apps from various classes
and the implications they can have on end-users.  All the analysis presented in
this section is performed using \generaldata.

\subsection{Market Staleness}
\label{sec:staleness}

\begin{figure*}
\centering
{\label{fig:ratings:boxplot:2012}{\includegraphics[width=\textwidth]{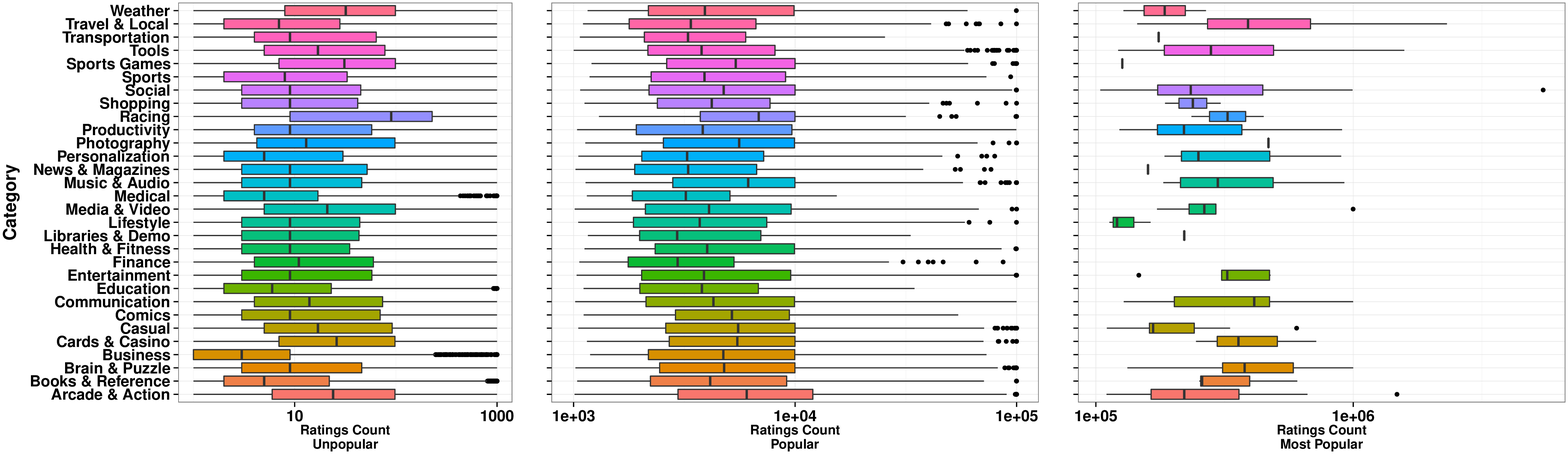}}}
{\label{fig:ratings:boxplot:14-15}{\includegraphics[width=\textwidth]{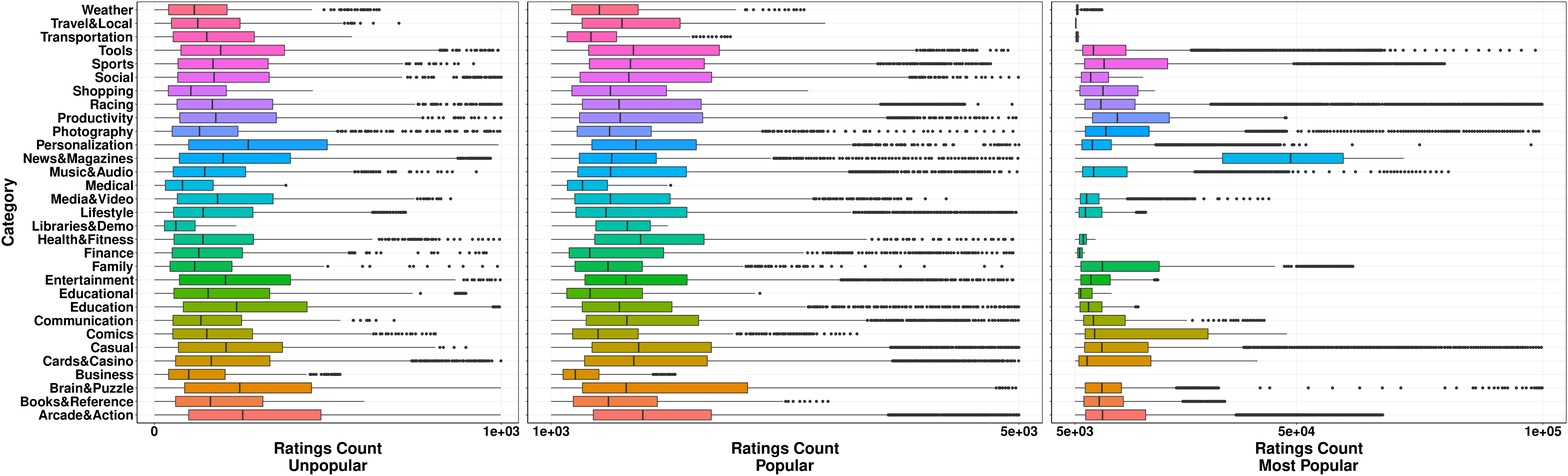}}}
\caption{Per app category distribution of rating counts.
(Top) {\bf Dataset.2012}. The
distribution is almost symmetric in the case of ``unpopular'' apps.  The
distributions for most of the categories are symmetric in the ``popular'' class
and span roughly from $1,000$ to $100K$ ratings.  The \textit{Business} and
\textit{Comics} categories do not have any apps in the ``most-popular'' class.
(Bottom) {\bf Dataset.14-15}. We observe smaller rating counts compared with
the apps in dataset.2012. This is natural, as these are new apps, thus likely
to receive fewer ratings. We also note that while a few ``Business'', ``Libraries \& Demo'' and ``Medical'' categories are unpopular and popular, none are
most popular.}
\label{fig:ratings}
\vspace{-15pt}
\end{figure*}

An important property of a market is its ``activity'', or how frequently
are apps being maintained. We say that an app is \textit{stale} if it has not
been updated within the last year from the observation period, and
\textit{active} otherwise.

The task of setting the app price is complex. However, relying on statistics
computed on the entire population, as opposed to only active apps, may mislead
developers. For instance, given that the listing price of apps forms a key
component of its valuation and sale, this becomes an important factor for fresh
developers trying to enter the market. Specifically, the median price in our
dataset is \$0.99 when all apps are considered and \$1.31 when considering only
active apps. This confirm our intuition that developers that set their price
based on the former value are likely to sell their apps at lower profits.

Figure~\ref{fig:agesince:2012} shows the box and whiskers
plot~\cite{benjamini1988opening} of the per-app time since the last update, by
app category, for dataset.2012. At most 50\% of the apps in each category have
received an update within a year from our observation period. For instance,
most apps in \textit{Libraries \& Demo} have not been updated within the last
1.5 years.  Some categories such as \textit{Arcade \& Action}, \textit{Casual},
\textit{Entertainment}, \textit{Books \& Reference}, \textit{Tools} contain
apps that are older than three years.

Figure~\ref{fig:agesince:14-15} plots this data for dataset.14-15.  Many
freshly uploaded apps were uploaded more recently: 50\% apps in each category
receive an update within 35 days, while apps in the ``Social'' and ``Tools''
categories received updates even within 15 days. This is natural, as new apps
may have more bugs and receive more developer attention.



Several reasons may explain the lack of updates received by many of the apps we
monitored. First, some apps are either stable or classic (time-insensitive apps,
not expected to change) and do not require an update.  Other
apps, e.g., e-books, wallpapers, libraries, do not require an update. 
Finally, many of the apps we monitored seemed to have been abandoned.

\subsection{App Popularity}
\label{sec:ratings}

\begin{table}
\centering
\textsf{
\begin{tabular}{l r r r}
\toprule
\textbf{Class} & \textbf{\# download} & \textbf{\% Dataset.2012} & \textbf{\% Dataset.14-15}\\
\midrule
Unpopular & $0$ -- $10^3$ & 74.14 & 77.55\\
Popular & $10^3$ -- $10^5$ & 24.1 & 18.43\\
Most-Popular & $> 10^5$ & 0.7 & 4.00\\
\bottomrule
\end{tabular}
}
\caption{
\normalfont{
Popularity classes of apps, along with their distribution.
Dataset.14-15 has a higher percentage of most-popular apps.}}
\label{table:rating-class}
\vspace{-15pt}
\end{table}

We propose to use the \textit{download count} to determine app popularity.
Higher rating counts mean higher popularity but not necessarily higher quality
(e.g., an app could attract many negative ratings).  Including unpopular apps
will likely affect statistics such as update frequencies: including unpopular
apps will lead to a seemingly counter-intuitive finding, indicating that most
apps do not receive any updates. Therefore, we classify apps according to their
popularity into three classes, ``unpopular'', ``popular'' and ``most-popular''.

Table~\ref{table:rating-class} shows the criteria for the 3 classes and the
distribution of the apps in dataset.2012 and dataset.14-15 in these classes.
The newly released apps have a higher percentage of unpopular apps, however,
surprisingly, they also have a higher percentage of ``most-popular'' apps. This
may be due to the fact that the newly released apps are more recent, coming at
a time of higher popularity of mobile app markets, and maturity of search rank
fraud markets (see $\S$~\ref{sec:discussion:threat}).

Figure~\ref{fig:ratings} (top) depicts the distribution of rating counts of
apps from dataset.2012, split by categories.  We observe that the
\textit{Business} and \textit{Comics} categories do not have any apps in the
\textit{Most-Popular} class, likely because of narrow audiences. From our data,
we observed that the median price of apps (\$1.99) in these categories is
significantly higher than the population (\$1.31) indicating lower competition.
The population in other categories is quite diverse with a number of outliers.
For instance, as expected, ``Angry Birds'' and ``Facebook'' are most popular
among the \textit{Most-Popular} class for \textit{Arcade \& Action} and
\textit{Social} categories, respectively. On the other hand, the distribution
is almost symmetric in case of \textit{Unpopular} except \textit{Business} and
\textit{Medical} categories where there are a number of outliers that are
significantly different from the rest of the population. We found that these
are trending apps --- apps that are gaining popularity. For instance, the free
app ``Lync 2010'' from ``Microsoft Corporation'' in \textit{Business} has 997
ratings. In case of \textit{Popular}, the distributions for most of the
categories are symmetric and span roughly from $1,000$ to $100K$ ratings where
75\% of apps have less than $10,000$ rating counts except \textit{Arcade \&
Action} category.

Figure~\ref{fig:ratings} (bottom) shows the same distribution for the apps
in dataset.14-15. We emphasize that the distribution is plotted over the
ratings counts at the end of the observation interval. Since these are newer
apps than those in dataset.2012, it is natural that they receive fewer ratings.
We also observe that several categories do not have apps that are in the ``most
popular'' category, including the ``Business'', ``Libraries \& Demo'' and
``Medical'' categories.

\subsection{App Updates}
\label{sec:updates}

\begin{figure*}
\centering
\subfigure[]
{\label{fig:updatehistogram:2012}{\includegraphics[width=0.32\textwidth]{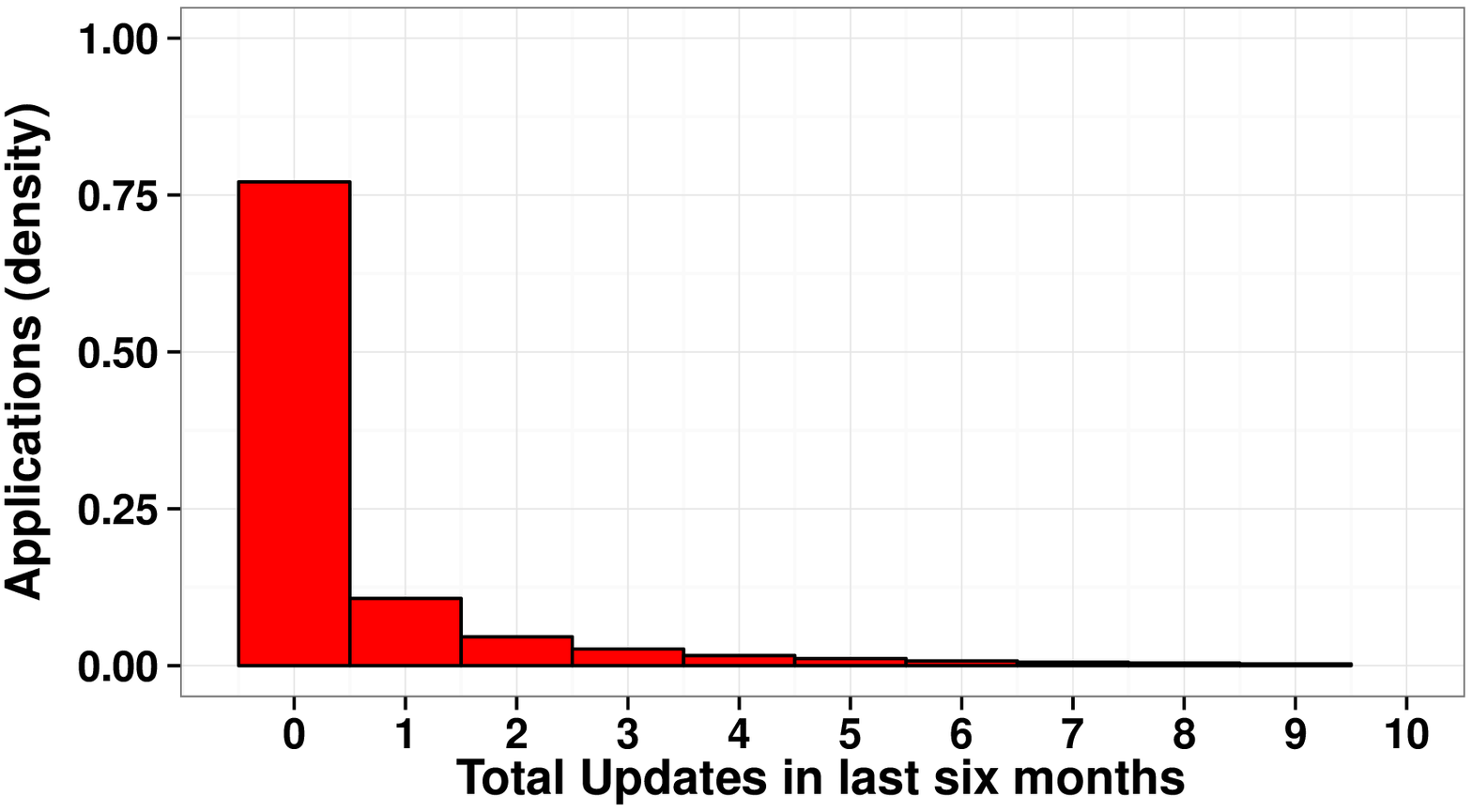}}}
\subfigure[]
{\label{fig:updatehistogram:14-15}{\includegraphics[width=0.32\textwidth, height=1.25in]{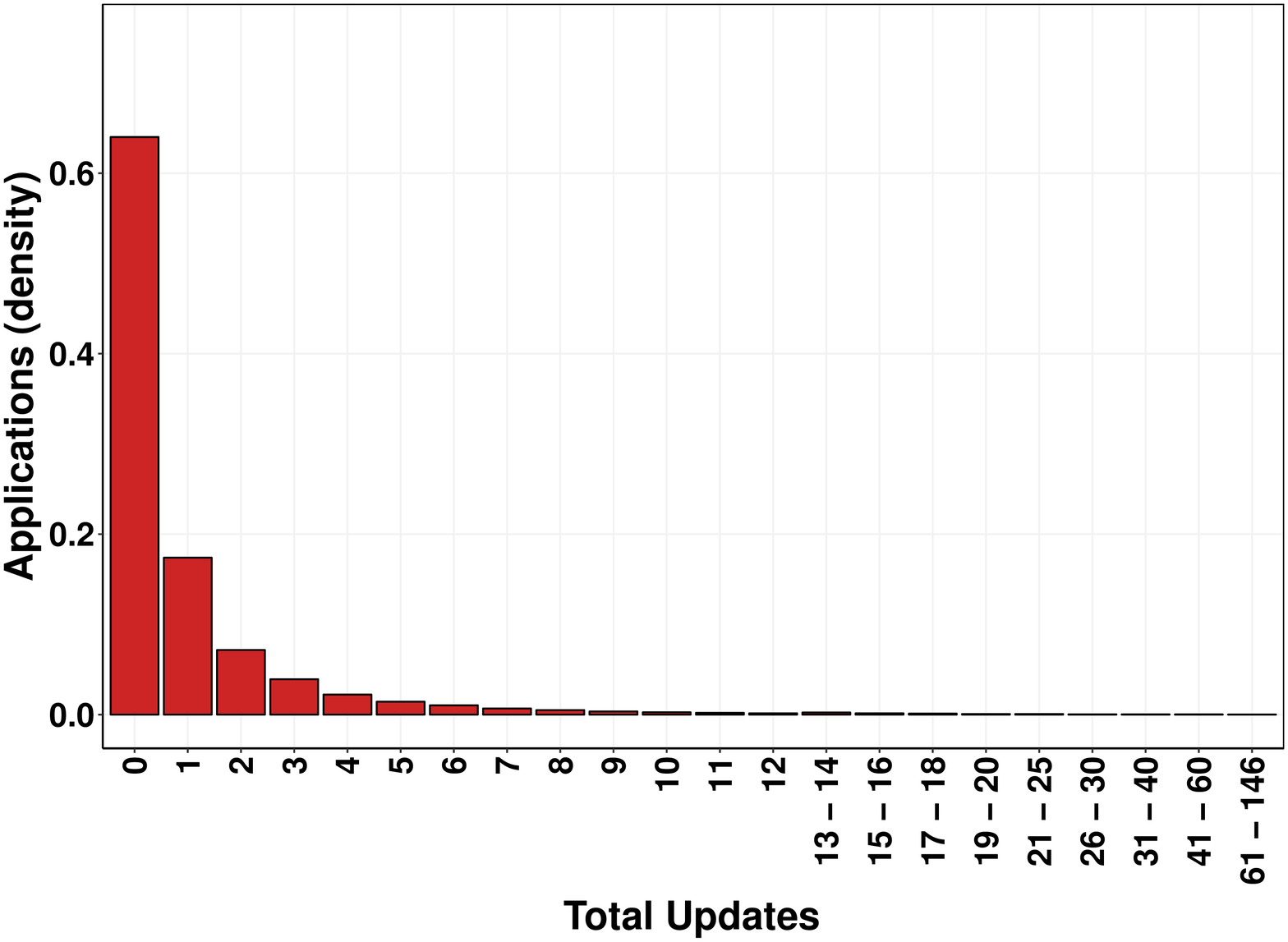}}}
\subfigure[]
{\label{fig:categorychangehistogram}{\includegraphics[width=0.32\textwidth,height=1.25in]{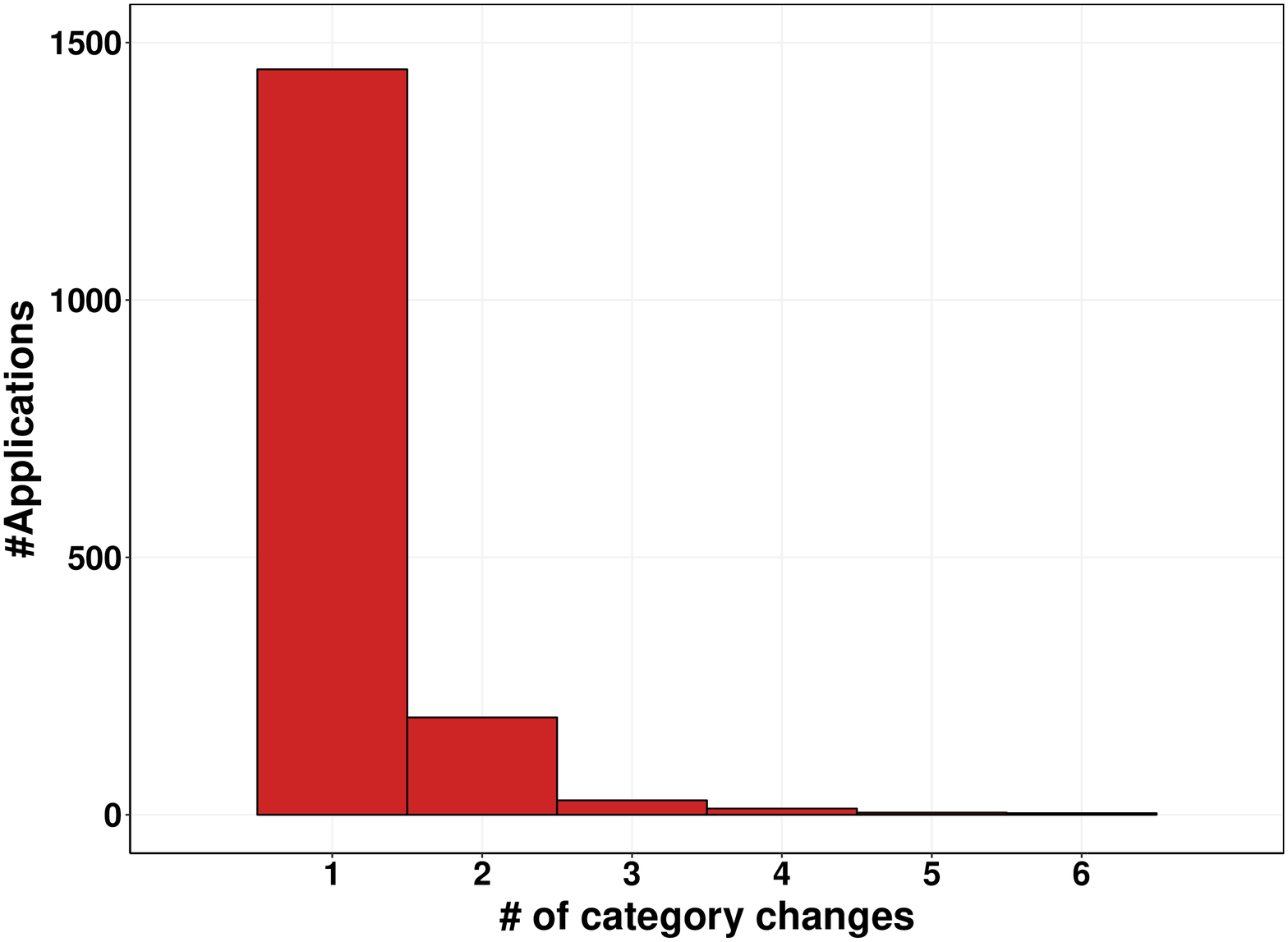}}}
\vspace{-10pt}
\caption{(a) Histogram of app updates for dataset.2012. Only 24\% apps have
received at least one update between April-November 2012.
(b) Histogram of fresh app updates for dataset.14-15. Unlike
the dataset.2012, 35\% of the fresh apps have received at least one update, while
1 app received 146 updates!
(c) Histogram of app category changes. 1.9\% apps have
received at least one category change between October 24, 2014 and May 5, 2015,
while several have received 6 category changes.
}
\vspace{-5pt}
\end{figure*}

Updates form a critical and often the last part of the software
lifecycle~\cite{ghezzi2002fundamentals}. We are interested in determining if
mobile app developers prefer seamless updating i.e., if they push out releases
within short time periods.

Fig.~\ref{fig:updatehistogram:2012} shows the distribution of the number of
updates received by the apps in dataset.2012. Only 24\% apps have received at
least one update within our observation period --- nearly 76\% have never been
updated. In contrast, Fig.~\ref{fig:updatehistogram:14-15} shows
that 35\% of the ``fresh'' apps in dataset.14-15 have received at least one
update within our observation period. Several apps received more than 100
updates, with one app receiving 146 updates in a 6 months interval. We
conjecture that this occurs because these are newly released apps, thus more
likely to have bugs, and to receive attention from their developers.

\begin{figure*}
\centering
{\label{fig:updatefrequency:2012}{\includegraphics[width=\textwidth]{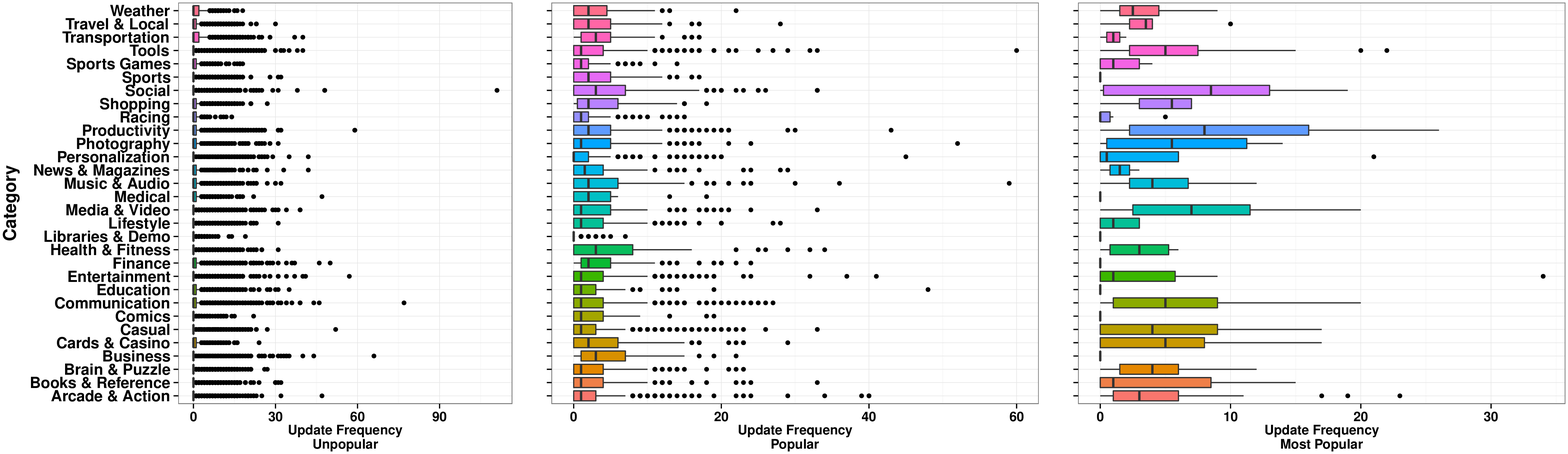}}}
{\label{fig:updatefrequency:14-15}{\includegraphics[width=\textwidth]{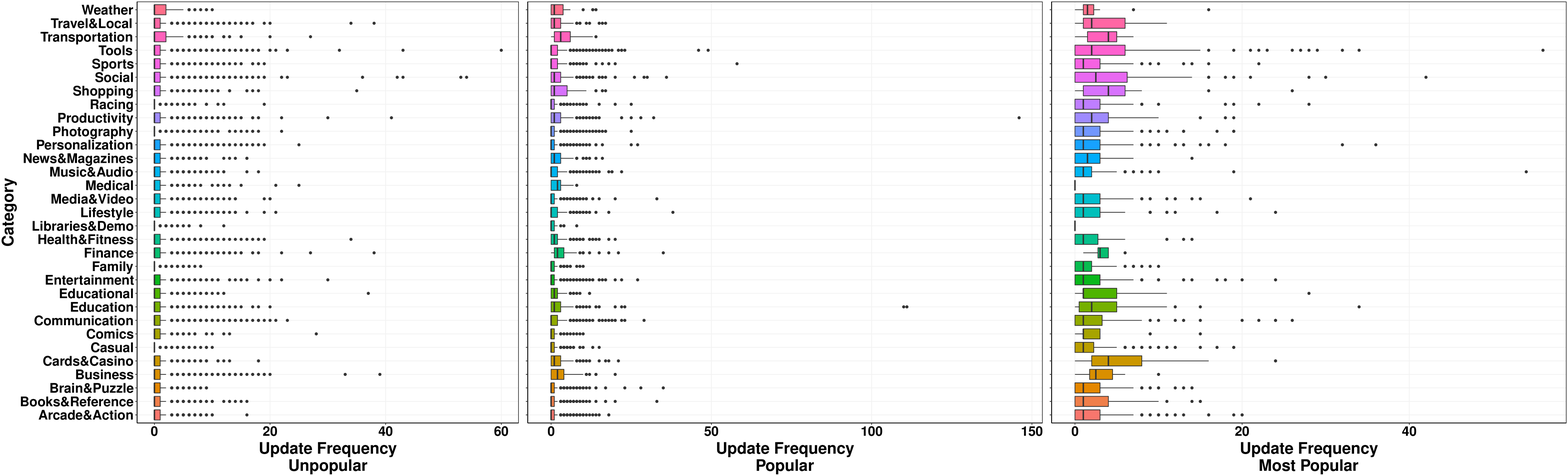}}}
\vspace{-10pt}
\caption{The distribution of update frequency, i.e., the update count for each
app per category.
(Top) {\bf Dataset.2012}. Unpopular apps receive few or no updates.  Popular apps
however received more updates than most-popular apps. This may be due to
most-popular apps being more stable, created by developers with well
established development and testing processes.
(Bottom) {\bf Dataset.14-15}. We observe a similar update count
distribution among unpopular apps to dataset.2012. Further, in the popular and
most popular classes, most app categories tend to receive fewer updates than
the dataset.2012 apps. However, a few apps receive significantly more updates,
with several popular apps receiving over 100 updates.}
\label{fig:update-freq}
\vspace{-15pt}
\end{figure*}

Figure~\ref{fig:update-freq} (top) plots the distribution of the update
frequency of the apps from dataset.2012, across categories based on their
popularity.  As expected, \textit{Unpopular} apps receive few or no updates. We
observed that this is due to the app being new or abandoned by its developer.
For instance, ``RoboShock'' from ``DevWilliams'' in \textit{Arcade \& Action}
with good reviews from 4 users has received only one update on September 28,
2012 since its release in August 2011 (inferred from its first comment).
Another app ``Shanju'' from ``sunjian'' in \textit{Social} has not been updated
since May 27, 2012 even though it received negative reviews. 

Outliers (e.g., ``Ctalk'' in the \textit{Social} category) push out large
number of updates ($111$). Popular apps are updated more frequently: 75\% in
each category receive 10 or fewer updates, while some apps average around 10-60
updates during our observation period.  User comments associated with these
apps indicate that the developer pushes out an update when the app attracts a
negative review (e.g., ``not working on my device!''). In the
\textit{Most-Popular} category, the population differs significantly. While
some apps seldom push any updates, apps like ``Facebook'' (\textit{Social})
have been updated 17 times.  The lower number of updates of most popular apps
may be due to testing: Companies that create very popular apps are more likely
to enforce strict testing and hence may not need as many updates as other apps.

To identify how frequently developers push these updates, we computed the
average update interval (AUI) per app measured in days (figure not shown). In
\textit{Popular} and \textit{Unpopular} classes, 50\% of apps receive at least
one update within 100 days. The most interesting set is a class of Unpopular
apps that receive an update in less than a week. For instance, the developer of
``Ctalk'' pushed, on average, one update per day totaling 111 updates in six
months indicating development stage (it had only 50-100 downloads) or
instability of the app.  On the other hand, \textit{Most-Popular} apps receive
an update within 20 to 60 days.

Figure~\ref{fig:update-freq} (bottom) shows the update frequency for the newly
released apps of dataset.14-15. Compared to the apps in dataset.2012, new
releases exhibit a similar update frequency distribution, with slightly lower
third quartiles. However, a few newly released popular apps receive
significantly more updates, some more than 100 updates.

\noindent
{\bf Updates, bandwidth and reputation.}
A high update frequency is a likely indicator of an on-going beta test of a
feature or an unstable application. Such apps have the potential to consume
large amounts of bandwidth.  For instance, a music player ``Player Dreams'',
with 500K-1M downloads, pushed out 91 updates in the last six months as part of
its beta testing phase (inferred from app description). With the application
size being around 1.8 MB, this app has pushed out $\approx$164 MB to each of
its users. Given its download count of 500K-1M, each update utilizes
$\approx$0.87-1.71 TB of bandwidth. We have observed that frequent
updates, especially when the app is unstable, may attract negative reviews.
For instance, ``Terremoti Italia'' that pushed out 34 updates in the
observation interval, often received negative reviews of updates disrupting the
workflow.

Furthermore, app market providers can use these indicators to
inform users about seemingly unstable applications and also as part of the
decision to garbage collect abandoned apps.


\subsection{App Category Changes}
\label{sec:category:change}

In the fresh app dataset.14-15 we found app category change events, e.g.,
``Social'' to ``Communication'', ``Photography'' to ``Entertainment'', between
different game subcategories. Such category changes may enable developers to
better position their apps and improve on their install and download count, as
categories may overlap, and apps may stretch over multiple categories.
Fig.~\ref{fig:categorychangehistogram} shows the distribution of the number of
app category changes recorded over the 6 months in dataset.14-15. Only 1.9\% of
apps have received at least one category change.
%

\section{Developer Impact}
\label{sec:developer:impact}

In this section, we are interested in understanding what fraction of popular
apps are being controlled by an elite set of developers and if there is a
power-law effect in-place. Next, we analyze the impact that developer actions
(e.g., changing the price, permissions etc.) can have on the app popularity.
We use dataset.2012 for this analysis.

\subsection{Market Control}

\begin{figure*}
\centering
\subfigure[]
{{\includegraphics[width=2.3in]{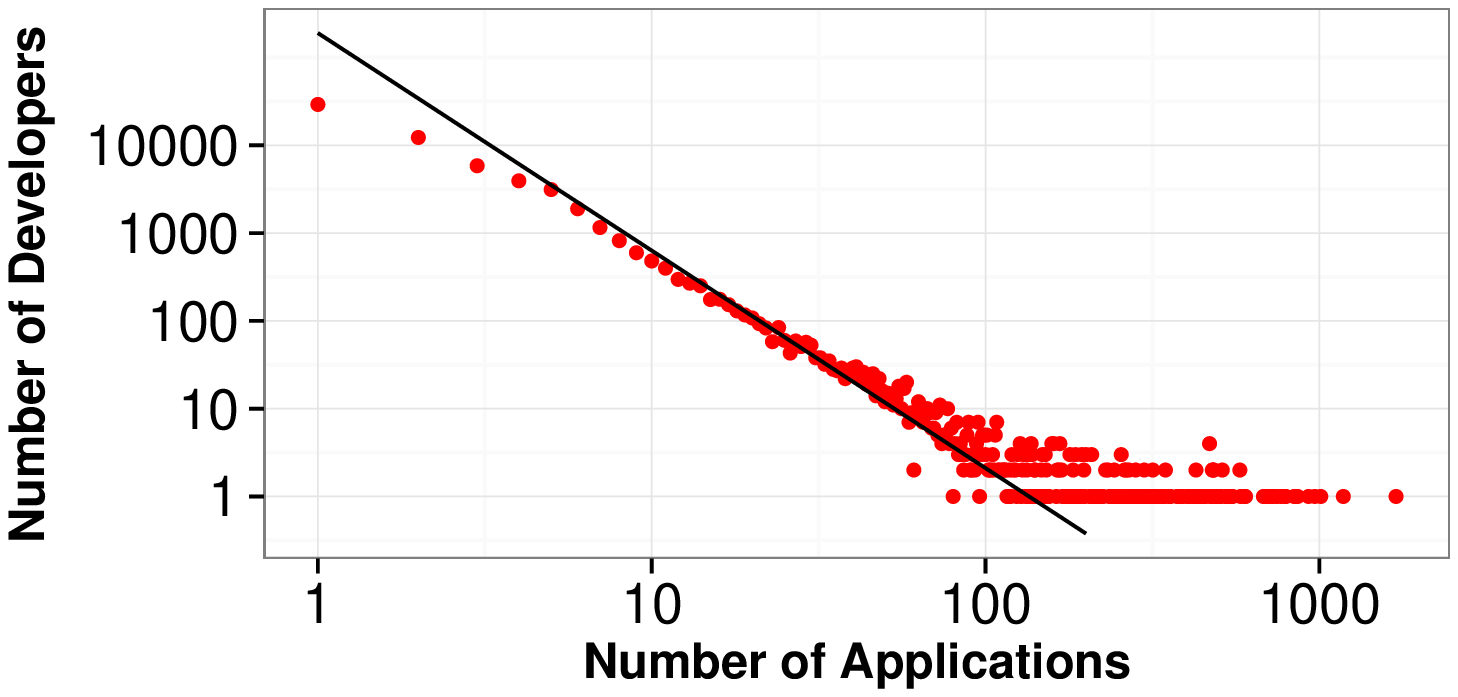}}}
\subfigure[]
{{\includegraphics[width=2.3in]{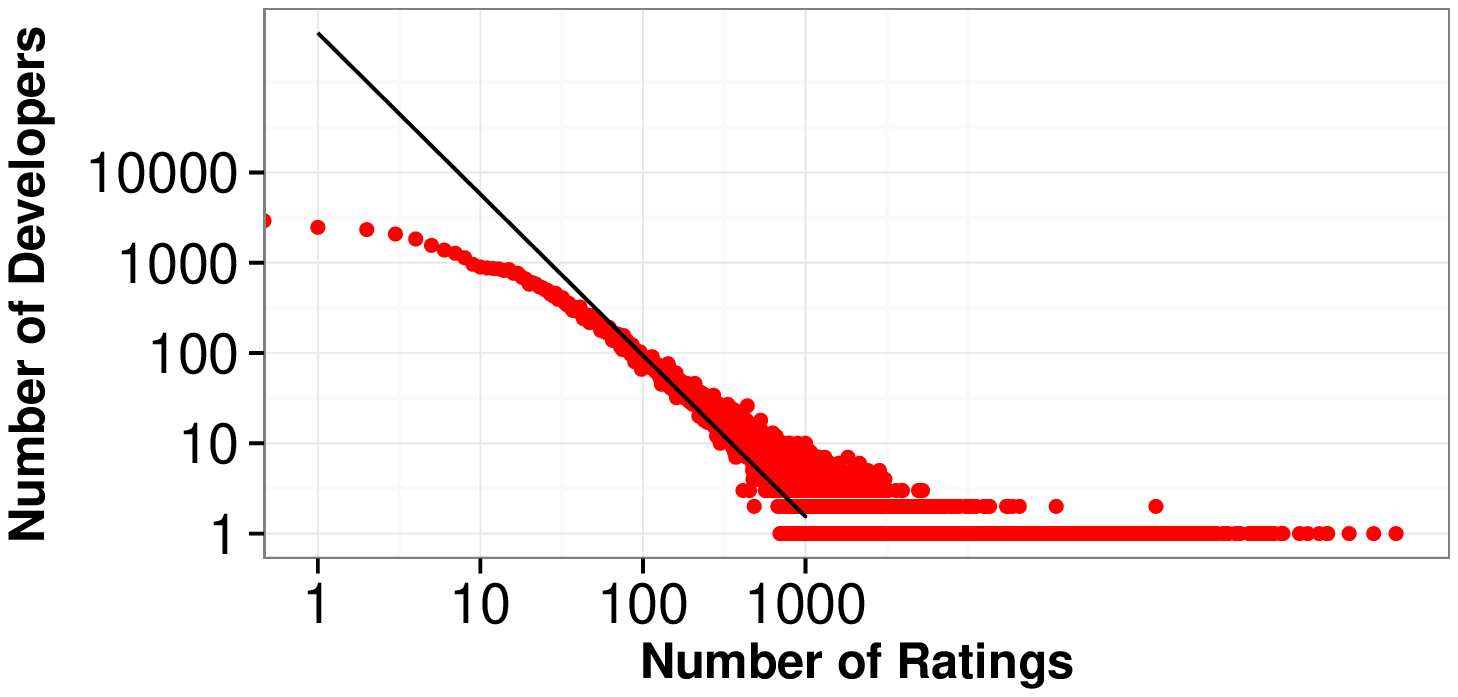}}}
\subfigure[]
{{\includegraphics[width=2.3in]{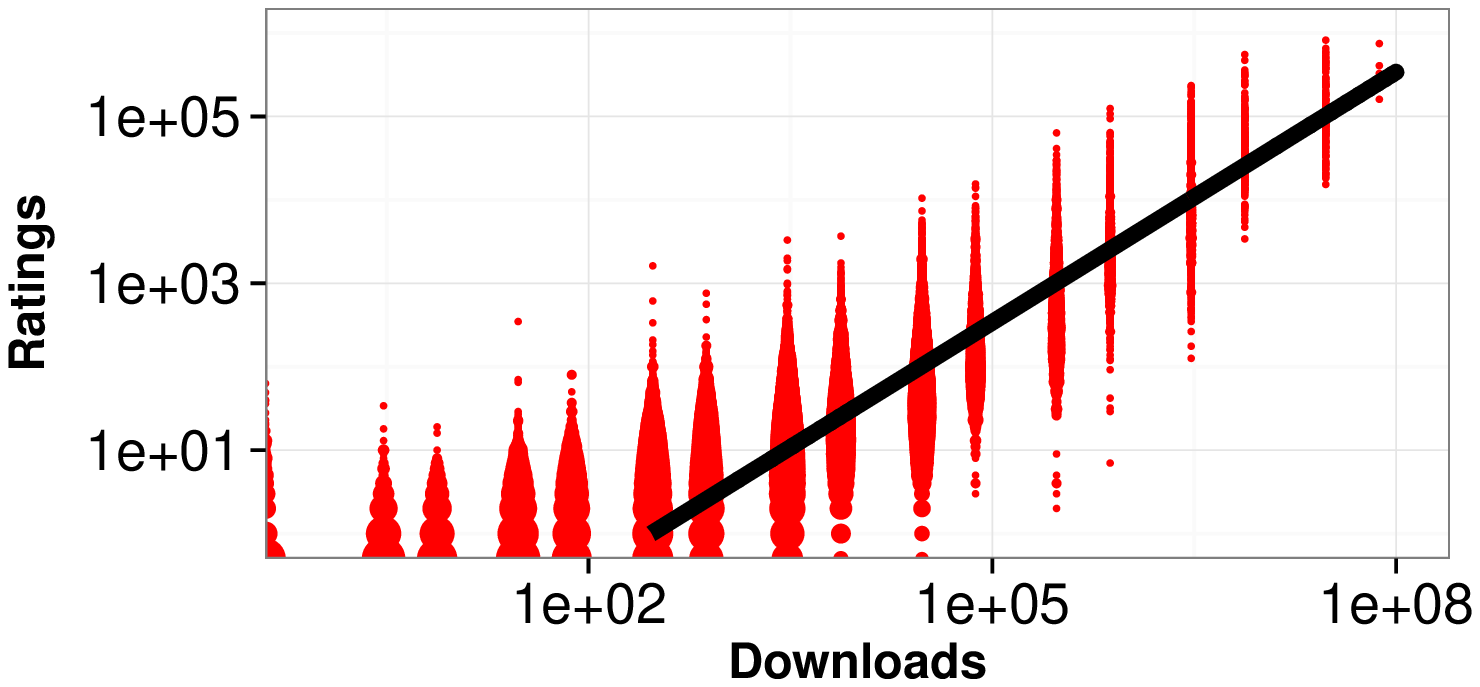}}}
\vspace{-10pt}
\caption{(a) Distribution of apps per developer. (b) Distribution of total
reviews per developer. (c) Scatter plot of downloads vs. ratings in Google
Play. Both axes are log-scaled. A linear curve was fitted with a slope of
0.00341 indicating that an application is rated once for about every 300
downloads.}
\label{fig:developer_app_fitting}
\end{figure*}

To understand the impact that developers have on the market, we observe their
number of apps, downloads, and review count.
Figure~\ref{fig:developer_app_fitting} plots these distributions, all showing
behavior consistent with a power-law distribution~\cite{mitzenmacher2004brief}.
We display the maximum likelihood fit of a power-law distribution for each
scatter plot as well~\cite{johnson2002applied,clauset2007power}.
%
%
Figure~\ref{fig:developer_app_fitting}(a) shows that a few developers have a
large number of apps while many developers have few apps.  However, the
developers that post the most apps do not have the most popular apps in terms
of reviews and download counts.  Instead,
Figure~\ref{fig:developer_app_fitting}(b) shows that a few developers control
apps that attract most of the reviews. Since
Figure~\ref{fig:developer_app_fitting}(c) shows an almost linear relation
between review and download counts (1 review for each 300 downloads), we
conclude that the apps developed by the controlling developers are popular. 


\subsection{Price Dispersion}
\label{sec:price}

\begin{figure}
\centering
\subfigure
{{\includegraphics[width=0.49\textwidth]{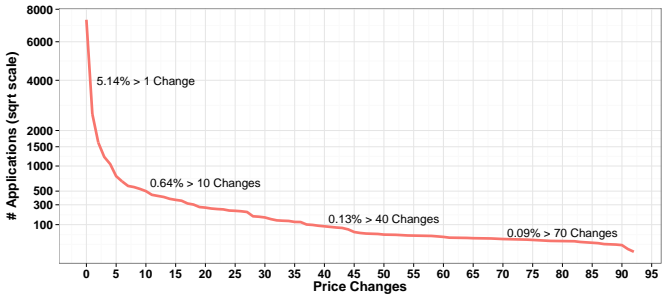}}}
\vspace{-5pt}
\caption{The (square root) of the number of apps whose number of
price changes exceeds the value on the $x$ axis. Only 5.14\% of the apps had a
price change, and 0.09\% of the apps had more than 70 changes.}
\label{fig:pricechanges}
\vspace{-5pt}
\end{figure}

\textit{Menu costs} (incurred by sellers  when making price changes) are lower
in electronic markets as physical markets incur product re-labeling
costs~\cite{levy1997magnitude}. In app markets menu costs are zero. We now
investigate if developers leverage this advantage i.e., if they adjust their
prices more finely or frequently.

Figure~\ref{fig:pricechanges} shows a variation of the complementary cumulative
distribution frequency (CCDF) of the number of price changes an app developer
made during our observation period. Instead of probabilities, the $y$ axis
shows the square root of the number of apps with a number of price changes
exceeding the value shown on the $x$ axis. We observe that 5.14\% of the apps
($\approx$4000) have changed their price at least once. The tail ($>$ 70
changes) is interesting --- about 23 apps are frequently changing their prices.
From our data, we observed that they are distributed as follows: \textit{Travel
\& Local} (11), \textit{Sports} (5), \textit{Business} (2), \textit{Brain \&
Puzzle} (2) and one in each of \textit{Education}, \textit{Finance}, and
\textit{Medical}. In this sample, ``LogMeIn Ignition'', developed by LogMeIn,
has 10K-50K downloads and underwent 83 price changes (Min:\$18.44, Max:\$27.80,
Avg:\$26.01, Stdev:\$2.01). The rest were either recently removed or are
unpopular.

\begin{figure}
\centering
\vspace{-15pt}
\subfigure
{{\includegraphics[width=0.49\textwidth]{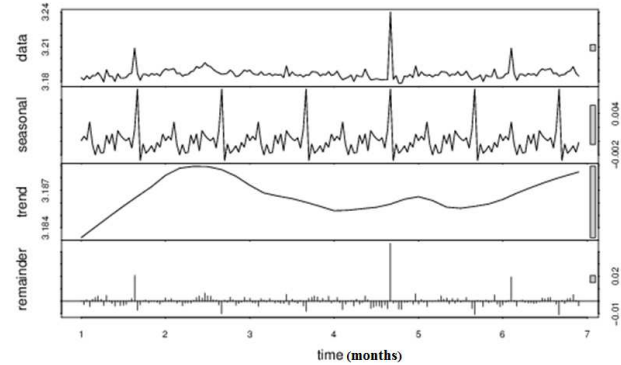}}}
\vspace{-15pt}
\caption{Monthly trend for the average app price. Over the 6 month observation
interval, the average app price does not exhibit a monthly trend.}
\label{fig:pricetimeline}
\vspace{-15pt}
\end{figure}

\textit{Price dispersion} is the spread between the highest and lowest prices
in the market.  In our dataset, we used the \textit{coefficient of variation}
(COV)~\cite{weber2004predicting}, the ratio of standard deviation to the mean,
to measure price dispersion. COV$=1$ indicates a dispersal consistent with a
Poisson process i.e., uniformly at random; COV$>1$ indicates greater
variability than would be expected with a Poisson process; and COV$<1$
indicates less variation. In our dataset, we observed an average COV (computed
for all apps) to be 2.45 indicating a non-negligible price dispersion, in
agreement with results in the context of other electronic
markets~\cite{brynjolfsson2000frictionless}.

Figure~\ref{fig:pricetimeline} shows the STL
decomposition~\cite{cleveland1990stl} of the average price time series in the
observation interval, for a periodicity of one month. The gray-bar on the
``monthly panel'' (see Figure~\ref{fig:pricetimeline}) is only slightly
larger than that on the ``data'' panel indicating that the monthly signal is
large relative to the variation in the data. In the ``trend'' panel, the gray
box is much larger than either of the ones on the ``data''/``monthly''
panels, indicating the variation attributed to the trend is much smaller than
the monthly component and consequently only a small part of the variation in
the data series. The variation attributed to the trend is considerably
smaller than the stochastic component (the remainders). We deduce that in our
six month observation period this data does not exhibit a trend.



\subsection{Impact of Developer Actions}
\label{sec:correlations}

Developers have control over several attributes they can leverage to increase
the popularity of their apps, e.g., pricing, the number of permissions
requested from users and the frequency of updates.  In this section we
investigate the relation between such levers and their impact on app
popularity.  For instance, common-sense dictates that a price reduction should
increase the number of downloads an app receives. 

\begin{table}
\centering
\begin{tabular}{|p{0.65cm}||p{0.6cm}|p{0.6cm}|p{0.6cm}|p{0.65cm}|p{0.6cm}|p{0.6cm}|p{0.6cm}|}
\hline 
 & {\scriptsize D $\uparrow$ }  & {\scriptsize P $\downarrow$ }  & {\scriptsize P $\uparrow$ }  & {\scriptsize RC $\uparrow$ }  & {\scriptsize SV $\uparrow$ }  & {\scriptsize TP $\downarrow$ }  & {\scriptsize TP $\uparrow$}\tabularnewline
\hline 
\hline 
{\scriptsize D $\uparrow$}  &  & {\scriptsize \cellcolor[gray]{0.909621}0.18} & {\scriptsize \cellcolor[gray]{1.000}-0.02} & {\scriptsize \cellcolor[gray]{0.931}0.13} & {\scriptsize \cellcolor[gray]{0.830}0.34} & {\scriptsize \cellcolor[gray]{0.954}0.09} & {\scriptsize \cellcolor[gray]{0.893}0.21}\tabularnewline
\hline 
{\scriptsize P $\downarrow$ }  & {\scriptsize \cellcolor[gray]{0.910}0.18 }  &  & {\scriptsize \cellcolor[gray]{1.000}-1.00} & {\scriptsize \cellcolor[gray]{0.954}0.09} & {\scriptsize \cellcolor[gray]{0.552}0.89} & {\scriptsize \cellcolor[gray]{0.553}0.89} & {\scriptsize \cellcolor[gray]{0.533}0.93}\tabularnewline
\hline 
{\scriptsize P $\uparrow$ }  & {\scriptsize \cellcolor[gray]{1.000}-0.02 }  & {\scriptsize \cellcolor[gray]{1.000}-1.00} &  & {\scriptsize \cellcolor[gray]{1.000}-0.23} & {\scriptsize \cellcolor[gray]{0.641}0.72} & {\scriptsize \cellcolor[gray]{0.743}0.51} & {\scriptsize \cellcolor[gray]{0.618}0.76}\tabularnewline
\hline 
{\scriptsize RC $\uparrow$ }  & {\scriptsize \cellcolor[gray]{0.931}0.13 }  & {\scriptsize \cellcolor[gray]{0.954}0.09} & {\scriptsize \cellcolor[gray]{1.000}-0.23} &  & {\scriptsize \cellcolor[gray]{0.636}0.73} & {\scriptsize \cellcolor[gray]{0.673}0.65} & {\scriptsize \cellcolor[gray]{0.648}0.70}\tabularnewline
\hline 
{\scriptsize SV $\uparrow$ }  & {\scriptsize \cellcolor[gray]{0.830}0.34 }  & {\scriptsize \cellcolor[gray]{0.552}0.89} & {\scriptsize \cellcolor[gray]{0.641}0.72} & {\scriptsize \cellcolor[gray]{0.636}0.73} &  & {\scriptsize \cellcolor[gray]{0.501}0.99} & {\scriptsize \cellcolor[gray]{0.500}1.00}\tabularnewline
\hline 
{\scriptsize TP $\downarrow$ }  & {\scriptsize \cellcolor[gray]{0.954}0.09 }  & {\scriptsize \cellcolor[gray]{0.553}0.89} & {\scriptsize \cellcolor[gray]{0.743}0.51} & {\scriptsize \cellcolor[gray]{0.673}0.65} & {\scriptsize \cellcolor[gray]{0.501}0.99} &  & {\scriptsize \cellcolor[gray]{1.000}-1.00}\tabularnewline
\hline 
{\scriptsize TP $\uparrow$ }  & {\scriptsize \cellcolor[gray]{0.893}0.21 }  & {\scriptsize \cellcolor[gray]{0.533}0.94} & {\scriptsize \cellcolor[gray]{0.618}0.76} & {\scriptsize \cellcolor[gray]{0.648}0.70} & {\scriptsize \cellcolor[gray]{0.500}1.00} & {\scriptsize \cellcolor[gray]{1.000}-1.00} & \tabularnewline
\hline 
\end{tabular}
\caption{
\normalfont{
Yule association measure for pairs of attributes for dataset.2012.
The sample size is the entire dataset for the observation
interval. D is number of downloads, P is price, RC is review count, SV is
software version number and TP is the total number of permissions.
($\uparrow$) denotes an increasing attribute and ($\downarrow$) denotes a
decreasing one.}
}
\label{tab:prob-table} %
\vspace{-15pt}
\end{table}

We study the association between app attribute changes. We define a random
variable for increase or decrease of each attribute, and measure the
association among pairs of variables. For example, let $X$ be a variable for
price increase. For each $\langle$ day, app $\rangle$ tuple, we let $X$ be a
set of all of the app and day tuples where the app increased its price that day
(relative to the previous day's value). For this analysis we consider 160K apps
that have changed throughout our observation period, and we discard the
remaining apps.  We use the Yule measure of association\cite{warrens2008} to
quantify the association between two attributes, $A$ and $B$:
$\frac{|A\cap B|*|\overline{A}\cap \overline{B}|-|A\cap \overline{B}|*|\overline{A}\cap B|}{|A\cap B|*|\overline{A}\cap \overline{B}|+|A\cap \overline{B}|*|\overline{A}\cap B|}$.


$\overline{A}$ is the complement of $A$, i.e., each $\langle$ day, app
$\rangle$ tuple where the attribute does not occur, and $|A|$ denote the
cardinality of a set (in this case $A$). This association measure captures the
association between the two attributes: zero indicates independence, +1
indicates perfectly positive association, and -1 perfectly negative
association.  Table~\ref{tab:prob-table} shows the measure values
for all pairs of download count (D), price (P), review count (RC) and total
number of permission (TP) attributes.

Table~\ref{tab:prob-table} shows that a price decrease has a high association
with changes in software version and permissions. However, similarly high
associations are not observed with a price increase. Thus, when a developer is
updating software or permissions they are more likely to decrease the price
than increase the price of an app.

We observed that changing the price does not show significant association with
the download or review counts. We randomly sampled 50 apps where this is
happening and observe the following to be the main reasons.  First, apps are
initially promoted as free and a paid version is released if they ever become
popular. However, in some cases, the feature additions are not significant
(e.g., ads vs. no ads) and hence do not cause enough motivation for users to
switch to the paid version. Second, with app markets offering paid apps for
free as part of special offers (e.g., Thanksgiving deals), users may expect the
app to be given out for free rather than take a discount of a few cents.





\section{Top-K Dynamics}
\label{sec:topk}

A higher position in Google's top-k lists (see $\S$\ref{sec:data:sets}) is
desirable and often attracts significant media attention~\cite{angrybirdstops}
which in turn increases the app popularity.
To analyze the dynamics of app in top-k lists, we have used the \topkdata
dataset (see $\S$\ref{sec:data:sets}). Google keeps the ranking algorithms for
the top-k lists secret. In this section we seek answers to several fundamental
questions: How long will an app remain on a top-k list?  Will an app's rank
increase any further than its current rank? How long will it take for an app's
rank to stabilize?


\subsection{Top-K App Evolution}
\label{sec:topk-movements}

\begin{table}
\textsf{
\begin{tabular}{|l | l |}
\hline
\textbf{Metric} & \textbf{Description}\tabularnewline
\hline
\hline
DEBUT & Debut rank (rank when it first gets onto the list)\tabularnewline
\hline
HRS2PEAK & Hours elapsed from debut until peak rank\tabularnewline
\hline
PEAK & Highest rank attained during its lifetime on the list\tabularnewline
\hline
TOTHRS & Total number of hours spent on the list\tabularnewline
\hline
EXIT & Exit rank (rank during the last hour on the list)\tabularnewline
\hline
RANKDYN & Total ranks occupied during its lifetime on the list\tabularnewline
\hline
\end{tabular}}
\caption{
\normalfont{
Scores proposed to study the evolution of apps on Top-K lists.}}
\label{tab:topkmetrics}
\vspace{-15pt}
\end{table}

\begin{figure*}
\centering
\subfigure[]
{\label{fig:topk_metrics_DEBUT}\includegraphics[width=0.32\textwidth]{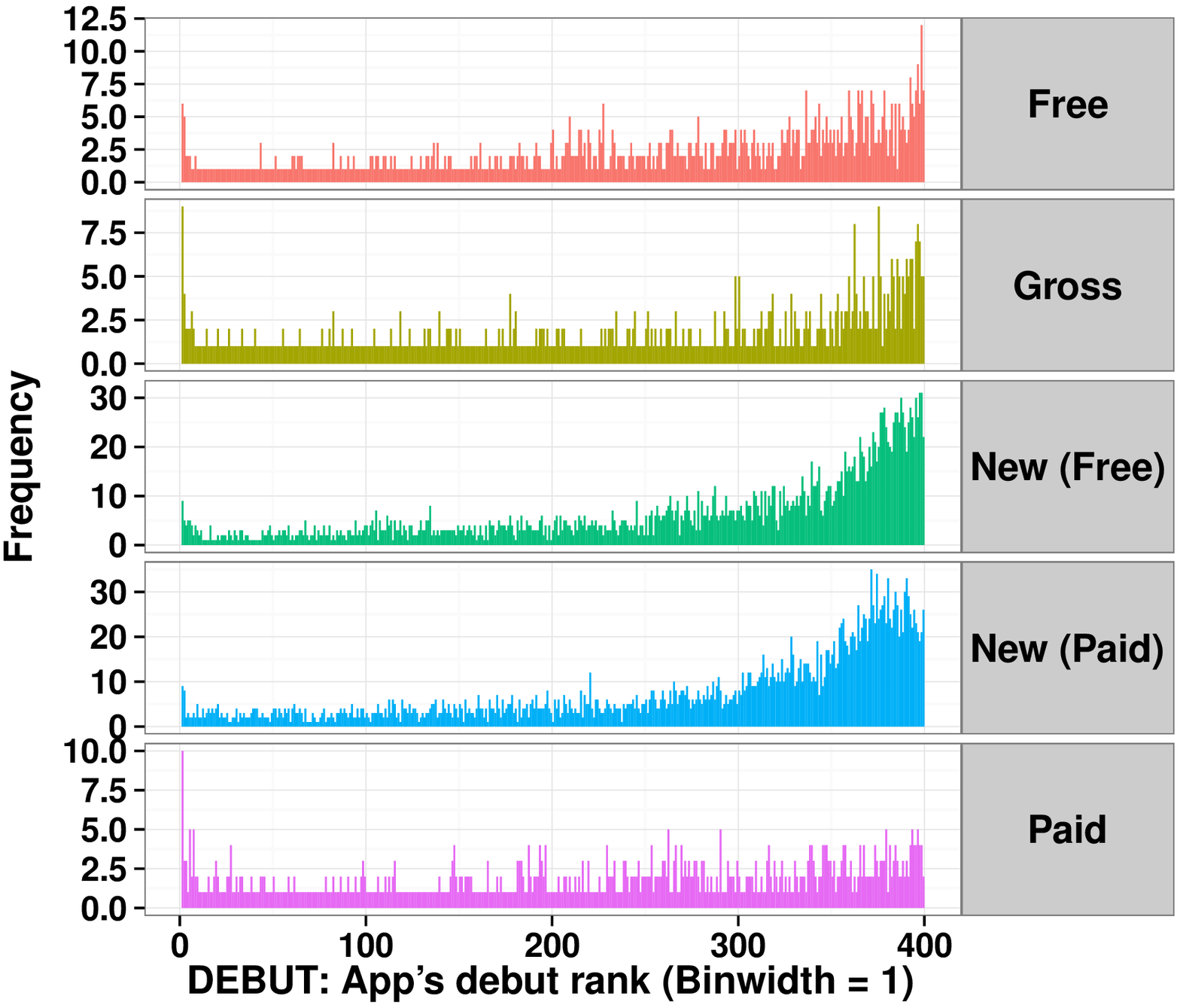}}
\subfigure[]
{\label{fig:topk_metrics_EXIT}\includegraphics[width=0.32\textwidth]{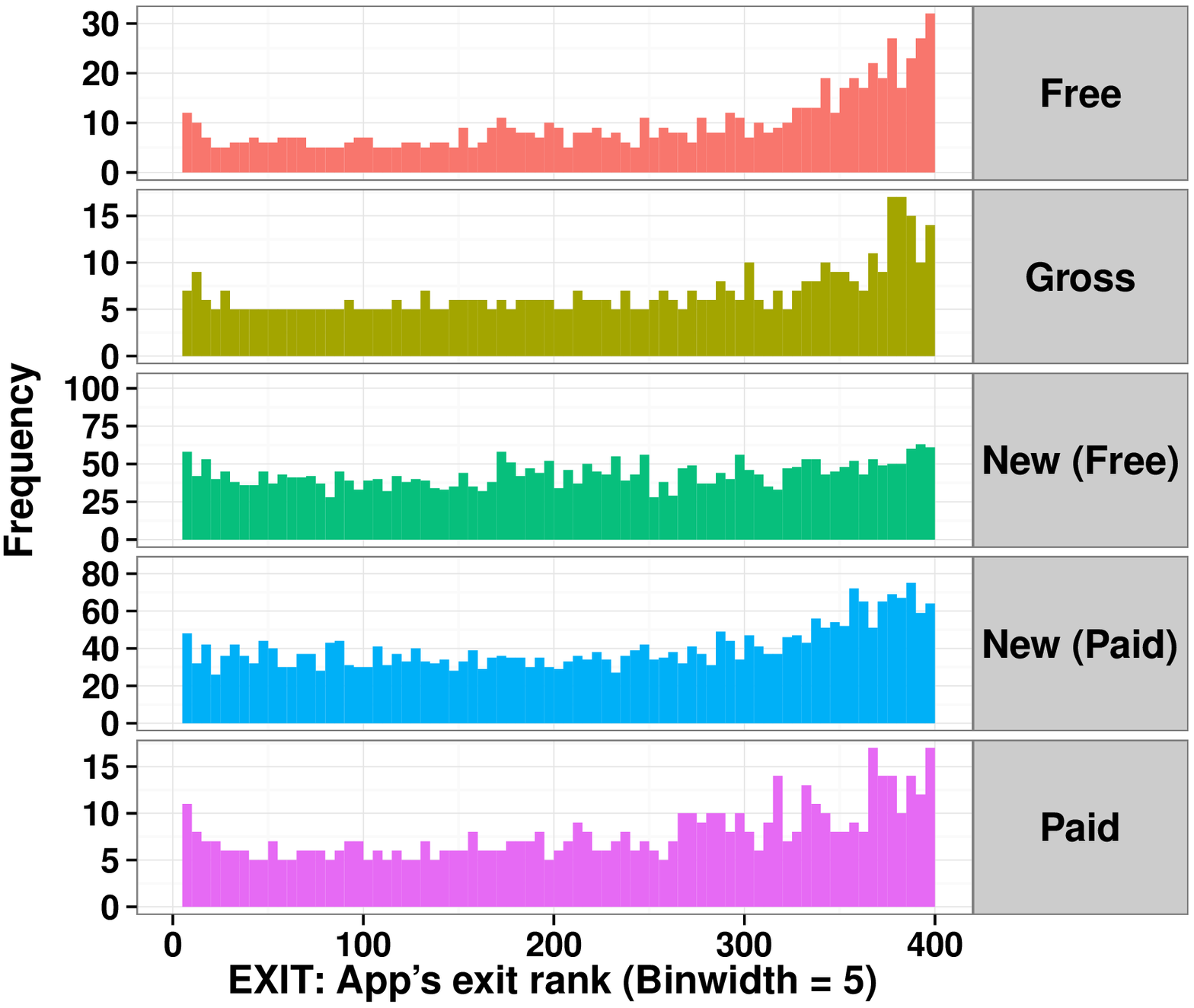}}
\subfigure[]
{\label{fig:topk_metrics_PEAK}\includegraphics[width=0.32\textwidth]{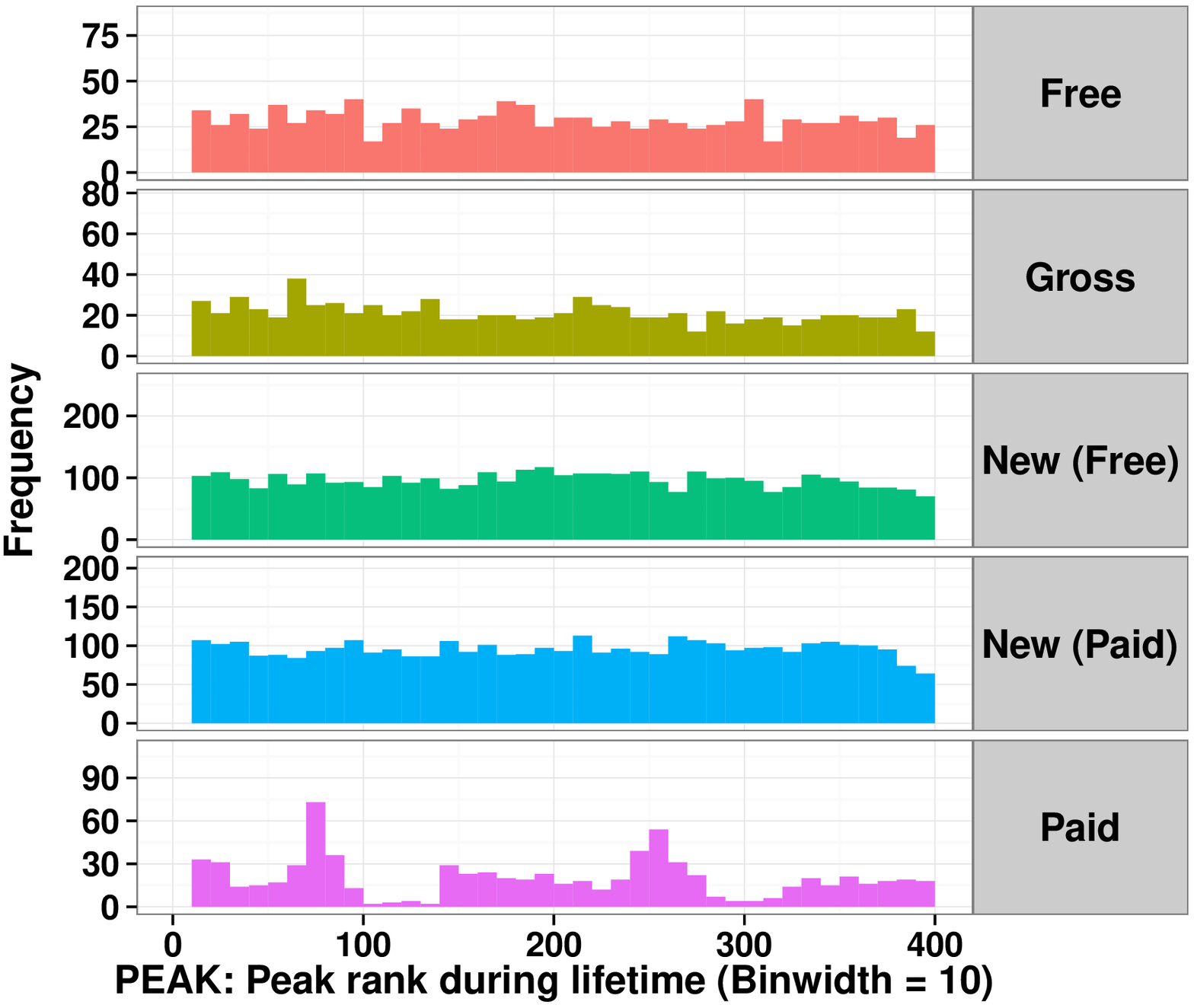}}
\caption{Distributions of (a) DEBUT, (b) EXIT and (c) achieved PEAK rank. The
$y$ axis is the number of apps whose ranks correspond to the values on the $x$
axis. Most apps entered and exited from the bottom of the list. The \textit{New
(Free)} and \textit{New (Paid)} lists choose apps updated within the last 20
days.}
\label{fig:topk:enter:exit}
\vspace{-15pt}
\end{figure*}

We investigate first whether apps follow the ``birth-growth-decline-death''
process (inverted bathtub curve~\cite{jiang2003aging}). Although every app's
path may be unique, we can summarize its life on a top-k list using the metrics
defined in Table~\ref{tab:topkmetrics}.

This six-tuple captures a suite of interesting information contained in each
app's list trajectory. To make these summaries comparable, we remove all
applications for which we are unable to compute the DEBUT information. For
instance, the set of applications obtained during the first hour of the crawl
process are removed.


Figures~\ref{fig:topk_metrics_DEBUT} and~\ref{fig:topk_metrics_EXIT} show the
histograms for the DEBUT and EXIT ranks, both indicating list positions, for
the 3000 apps we monitored. Smaller numbers indicate better performance. The
plots show that most apps entered and exited from the bottom part of the list
(indicated by the high debut and exit ranks). This is consistent with the
lifetime metaphor discussed earlier. However, a small number of apps entered
the list highly ranked. For instance, in the \textit{Paid}
category, the best DEBUT was attained by ``ROM Manager'', by ``Koushik Datta''
that entered at \#1 on August 14, 2012, and exited at rank \#20 on October 6,
2012, occupying seven different ranks during its lifetime on the list.  Another
noteworthy DEBUT was attained by ``Draw Something'', by ``OMGPOP'' that entered
at \#2 on October 1, 2012, peaked to \#1 on Oct 25, 2012 and exited at Nov 6,
2012 at \#4. During its lifetime, the worst rank it achieved was \#38.

\begin{figure*}
\centering
\subfigure[]
{\label{fig:topk_metrics_HRS2PEAK}\includegraphics[width=0.32\textwidth]{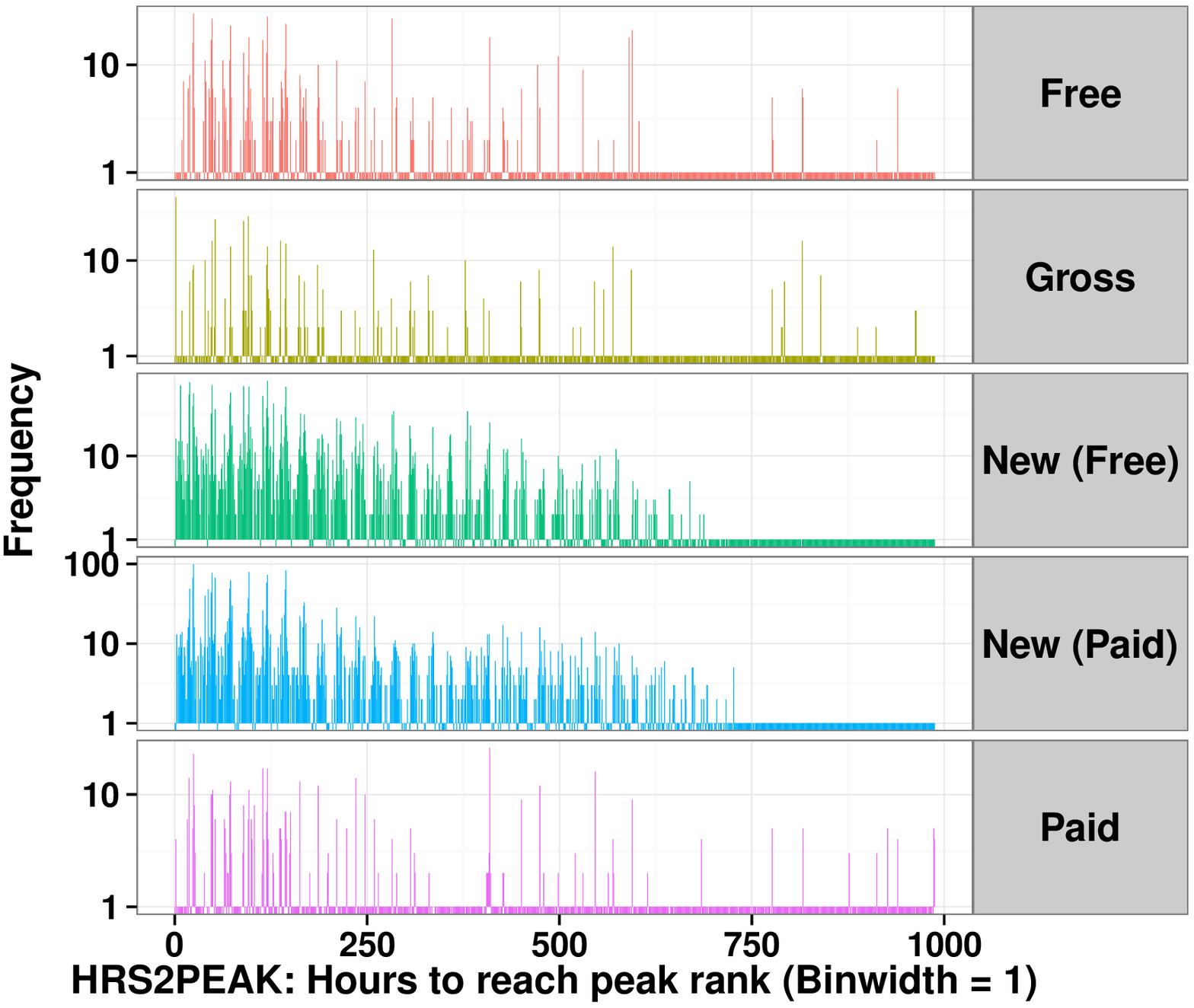}}
\subfigure[]
{\label{fig:topk_metrics_TOTHRS}\includegraphics[width=0.32\textwidth]{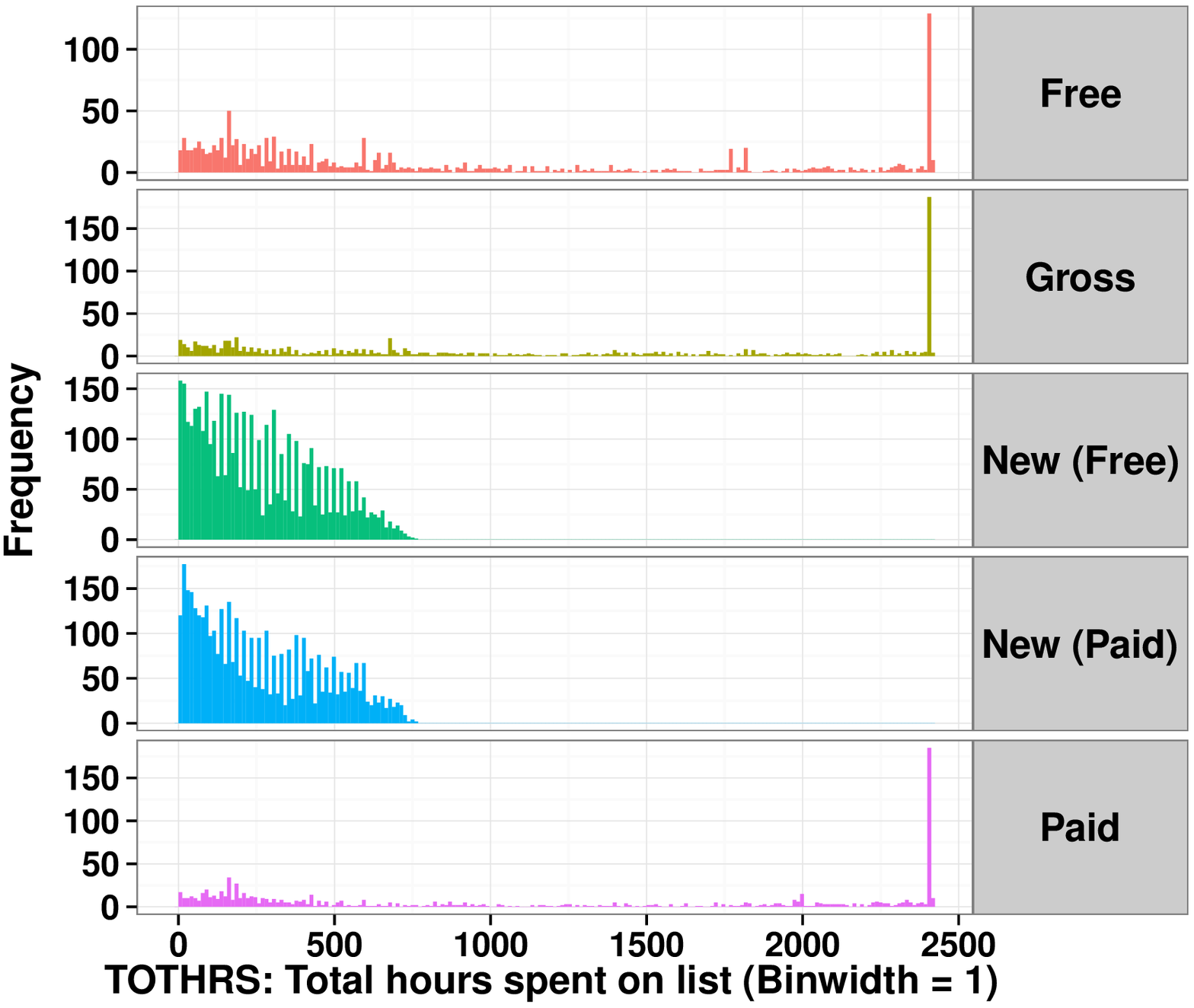}}
\subfigure[]
{\label{fig:topk_metrics_RANKDYN}\includegraphics[width=0.32\textwidth]{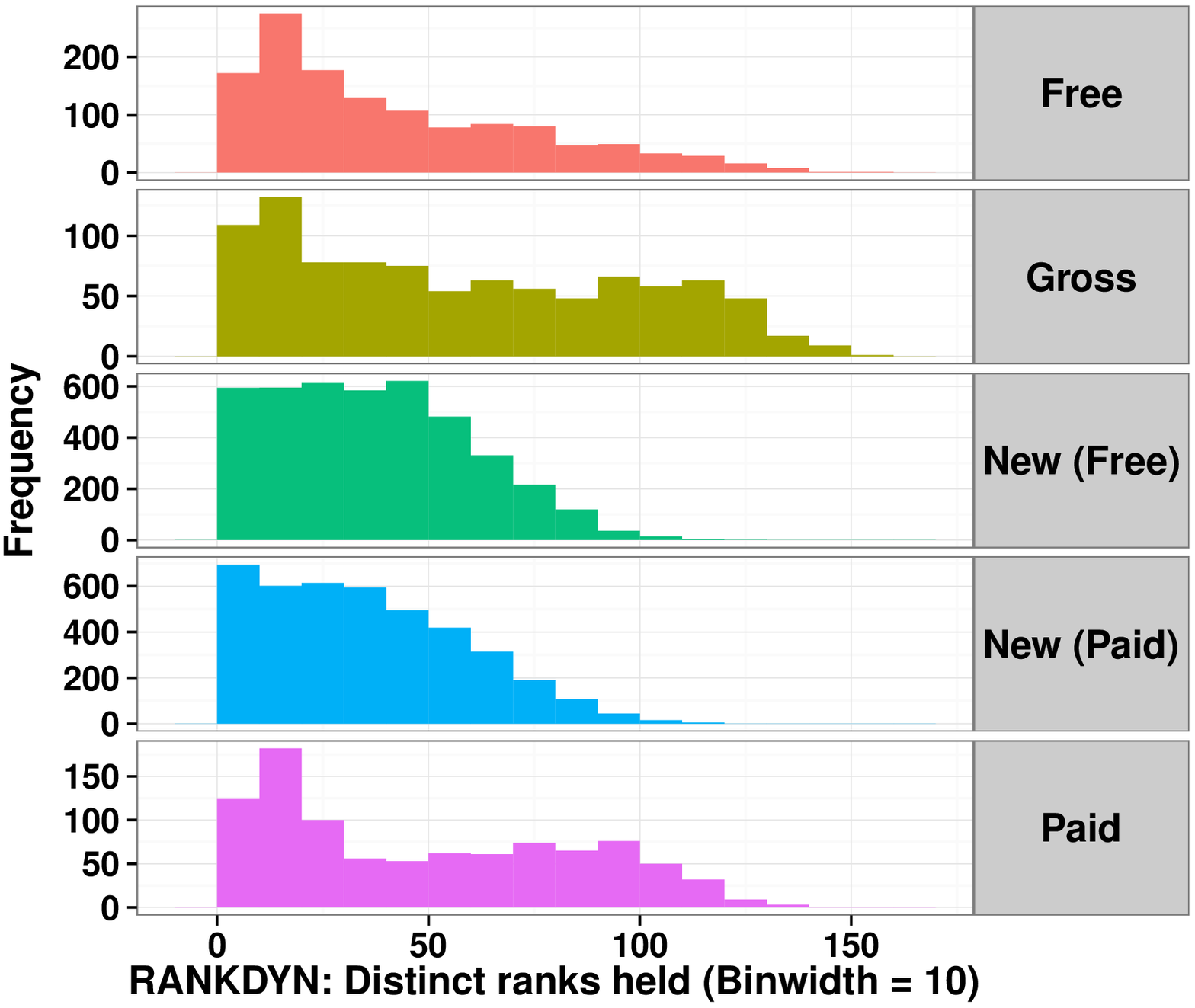}}
\caption{Distributions of (a) required HRS2PEAK, (b) spent TOTHRS
and (c) RANKDYN over lifetime. The $y$ axis displays the number of apps whose
hours correspond to the values displayed on the $x$ axis. \textit{New (Free)}
and \textit{New (Paid)} apps do not stay on the list for more than 500 hours.
While few \textit{New (Free)} and \textit{New (Paid)} apps achieve more than
100 different ranks, \textit{Gross} list apps achieve an almost uniform
distribution up to 125 different ranks.}
\label{fig:topk:hours}
\vspace{-15pt}
\end{figure*}

Figure~\ref{fig:topk_metrics_PEAK} shows the distribution of the peak rank
achieved in top-k lists (PEAK) and Figure~\ref{fig:topk_metrics_HRS2PEAK} shows
the distribution of the number of hours required for apps to reach the peak
(HRS2PEAK).

Figure~\ref{fig:topk_metrics_TOTHRS} shows the total number of hours spent by
apps in the top-k lists (TOTHRS). Figure~\ref{fig:topk_metrics_RANKDYN} shows
the number of ranks achieved in a top-k list (RANKDYN): very few \textit{New
(Free)} and \textit{New (Paid)} apps achieve more than 100 different ranks,
with most apps achieving 50 or fewer. This differs significantly in the other
top-k lists. 

Among the many apps with poor DEBUT and EXIT positions, most had a short,
uneventful life (i.e., low TOTHRS, poor PEAK, low HRS2PEAK), but several were
able to reach a high peak position and/or remain for a long time. One entry,
``PS Touch'', by ``Adobe'', entered at \#413 and has been on the ``Gross'' list
for 2,403 hours ($\approx3$ months, although it peaked only at \#206. This app
also took a remarkably slow journey (more than two months) to reach that peak
and has occupied 137 distinct ranks. Also note that 67 apps attained their top
rankings in their debut hour (i.e., HRS2PEAK = 1). Many of these apps stayed on
the list for a very short time, but there are 6 apps that stayed for 100 - 1000
hours, 31 stayed for more than 1000 hours.

Figure~\ref{fig:topk_metrics_TOTHRS} shows that \textit{New (Free)} and
\textit{New (Paid)} apps do not stay on the list for more than 500 hours
($\approx20$ days) indicating that these lists may be taking into account all
those applications which were last updated in the last 20 days. We have
confirmed this hypothesis also by verifying that indeed the ``last updated''
field of these apps is within the last 20 days.  From the same figure, for
other lists, we also emphasize the presence of a long tail of apps that have
been present for thousands of hours. We conclude that: (1) a
majority of apps follows a ``birth-growth-decline-death'' process, as they
enter/exit from the bottom part of a list, (2) most of the apps with modest
DEBUT and EXIT values have a short, eventful life occupying many ranks quickly,
and (3) the \textit{New (Free)} and \textit{New (Paid)} lists choose among apps
that were updated within the last 20 days.

\subsection{Top-K List Variation}
\label{sec:topk-temporal}

We now characterize the changes in the rankings of the top-k items from the
five lists over time.


We use the \textit{Inverse Rank Measure} to assess the changes over time in
each of the rankings. This measure gives more weight to identical or near
identical rankings among the top ranking items. This measure tries to capture
the intuition that identical or near identical rankings among the top items
indicate greater similarity between the rankings.  Let us assume the following:
$k_n$ is the list of top-k apps at time $t_n$, $\sigma_n(i)$ is the rank of app $i$
in $k_n$, $Z$ is the set of items common to $k_{n-1}$ and $k_n$, $S$ is the set
of items in $k_{n-1}$ but not in $k_{n}$, $T$ is the set of items in $k_{n}$
but not in $k_{n - 1}$. Then, the inverse rank measure is~\cite{bar2007some}
defined as $M^{(k_{n-1}, k_{n})} = 1 - \frac{N^{(k_{n-1}, k_{n})}}{Nmax^{(k_{n-1}, k_{n})}}$, where $N^{(k_{n-1},k_{n})} = 
\sum_{i \in Z}|\frac{1}{\sigma_{n-1}(i)} - \frac{1}{\sigma_{n}(i)}|+ $ $\sum_{i \in S}|\frac{1}{\sigma_{n-1}(i)} - \frac{1}{(|k_{n}|+1)}| + \sum_{i \in T}|\frac{1}{\sigma_{n}(i)} - \frac{1}{(|k_{n-1}|+1)}|$, and
$Nmax^{(k_{n-1},k_{n})} = \sum_{i = 1}^{|k_{n-1}|}|\frac{1}{i} - \frac{1}{(|k_{n}| + 1)}| + \sum_{i = 1}^{|k_{n}|} |\frac{1}{i} - \frac{1}{(|k_{n-1}| + 1)}|$.

\begin{figure*}
 \centering
   \subfigure[]
    {\label{fig:topk_M-topk_inverse}\includegraphics[width=0.3\textwidth]{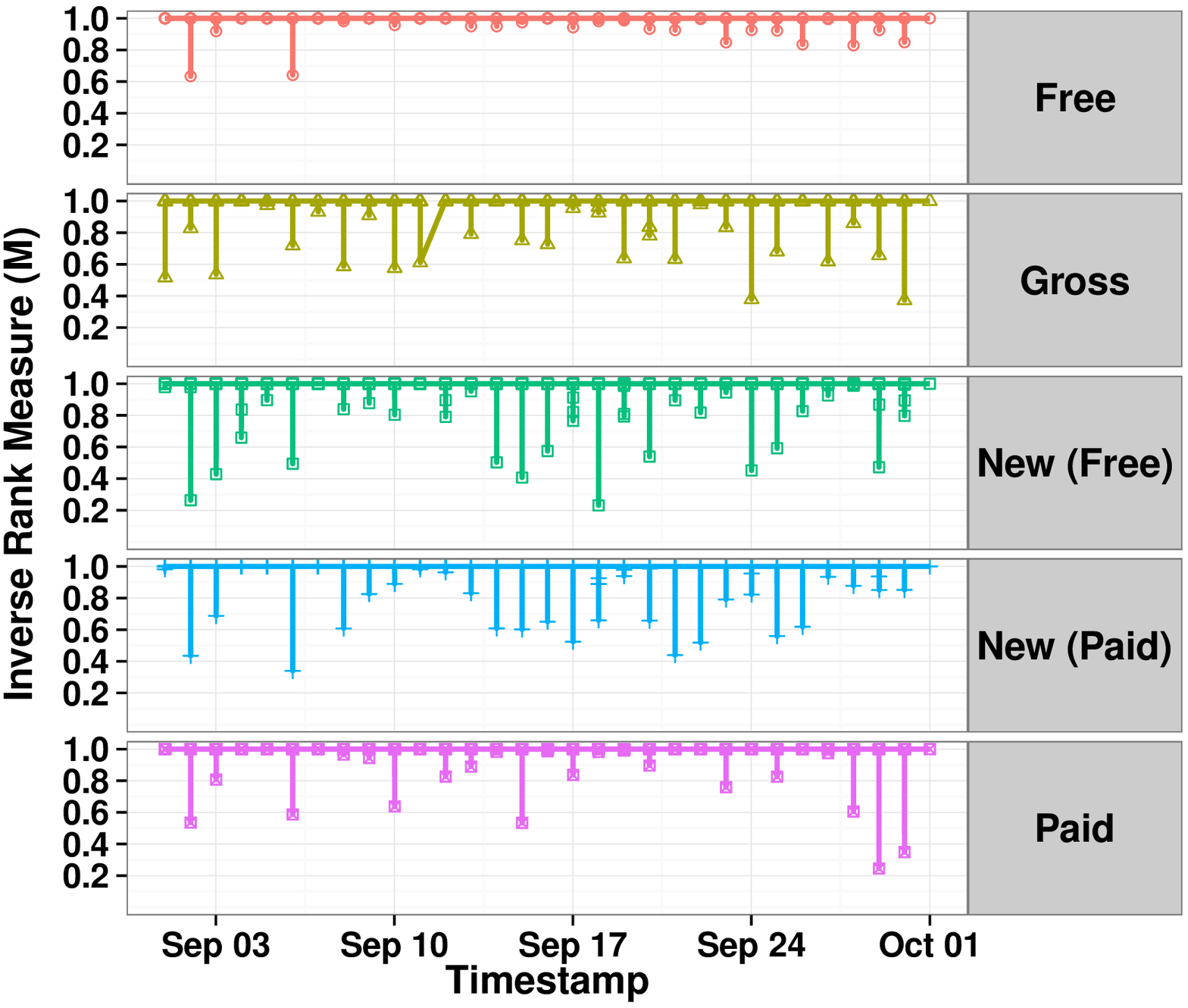}}
   \subfigure[]
    {\label{fig:topk_M-topk_occupants}\includegraphics[width=0.3\textwidth]{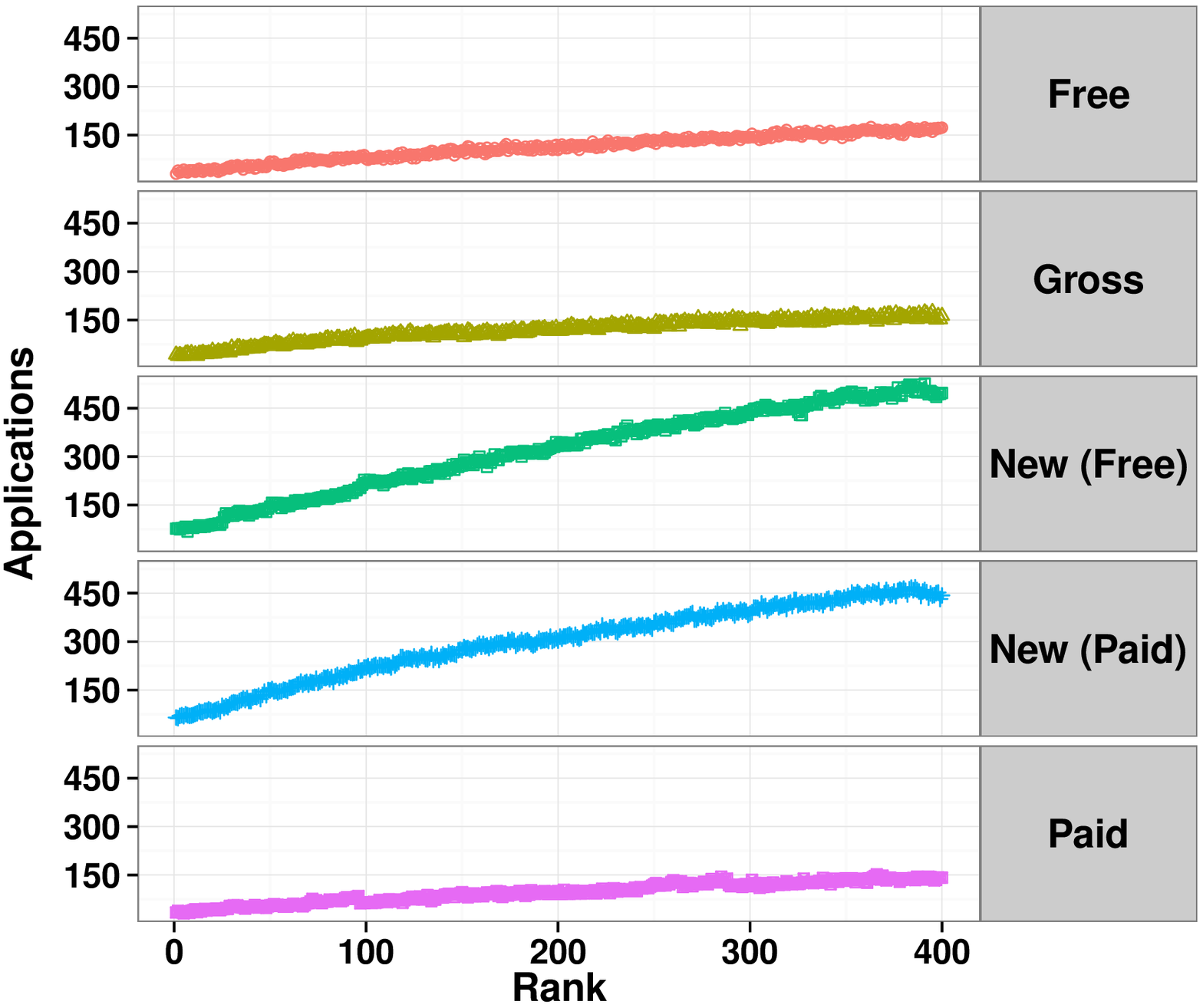}}
   \subfigure[]
    {\label{fig:topk_M-topk_lifetime_densityplot}\includegraphics[width=0.33\textwidth]{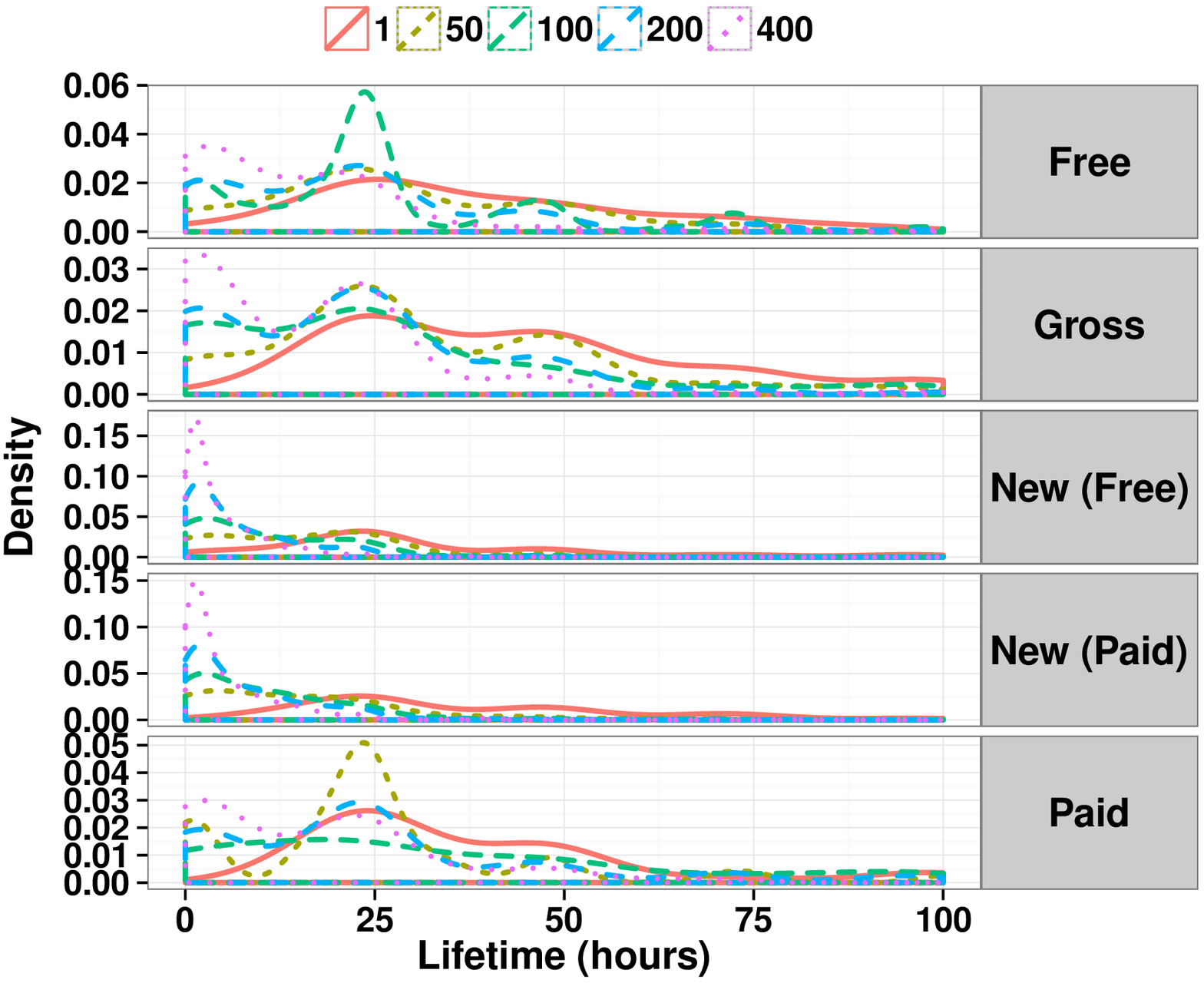}}
\caption{(a) The Inverse Rank Measure vs. Timestamp. The \textit{Free} list
varies little from day to day, which is not the case for \textit{Paid} and
\textit{Gross}.
(b) Number of apps vs. ranks.
(c) Lifetime of apps at various ranks. The average top-k list lifetime is
longer for higher ranking than for lower ranking apps.}
\label{fig:topk_M}
\end{figure*}

Figure~\ref{fig:topk_M-topk_inverse} shows the variation of
$M^{k_{t_1},k_{t_2}}$ for consecutive days in the month of September. Note that
values above 0.7 indicate high similarity~\cite{bar2007some}.  We observe that
the lists are similar from day to day for \textit{Free} list but this is not
the case for \textit{Paid} and \textit{Gross}. Intuitively, this indicates that
the effort to displace a free app seems to be higher than that of a paid app or
the frequency with which the ranking algorithm is run on \textit{Free} list is
less than that of the \textit{Paid} list.

\begin{figure*}
\centering
\subfigure[]
{\label{fig:timeline_pos_reviews}\includegraphics[width=0.32\textwidth]{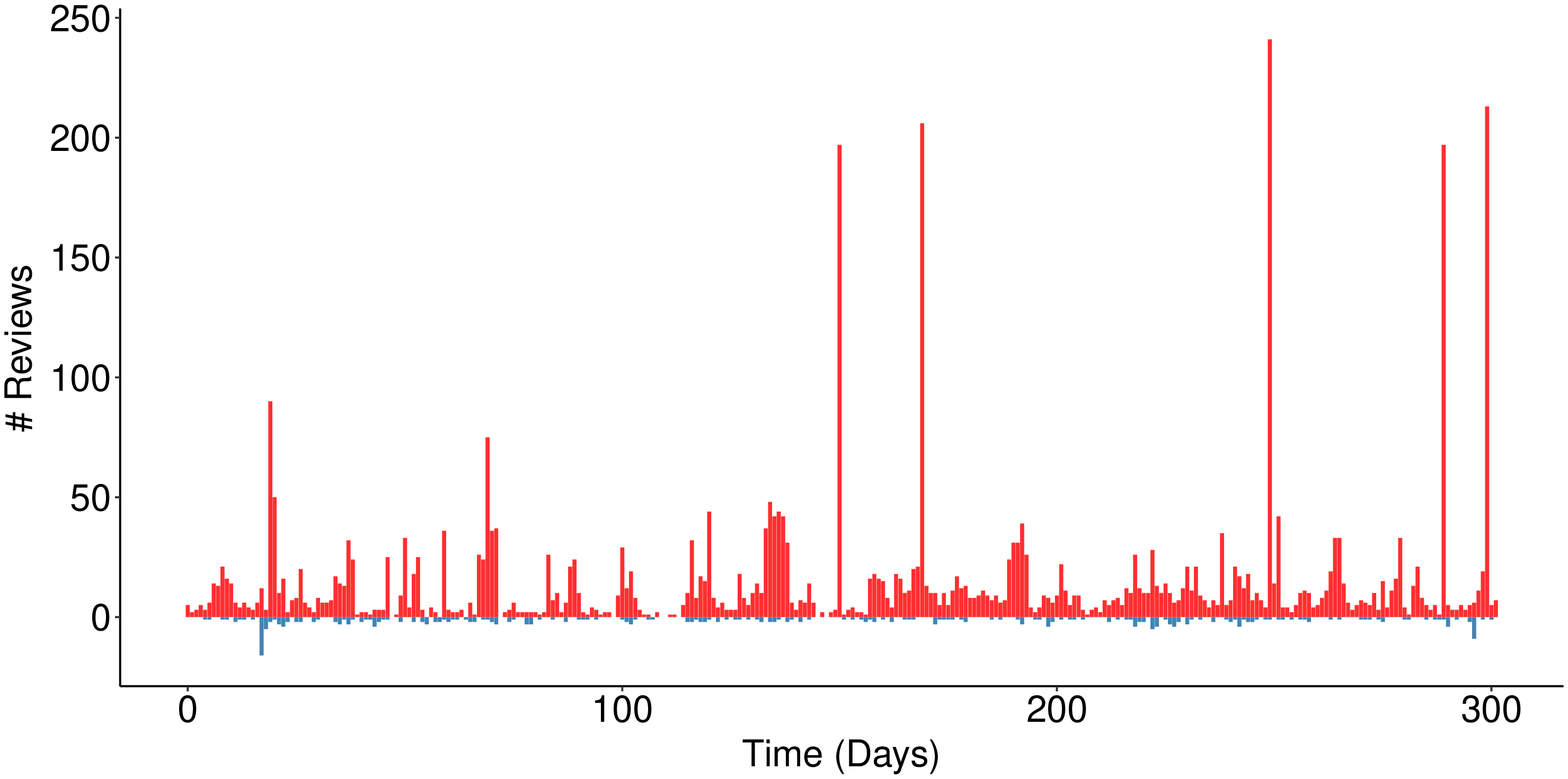}}
\subfigure[]
{\label{fig:timeline_neg_reviews}\includegraphics[width=0.32\textwidth]{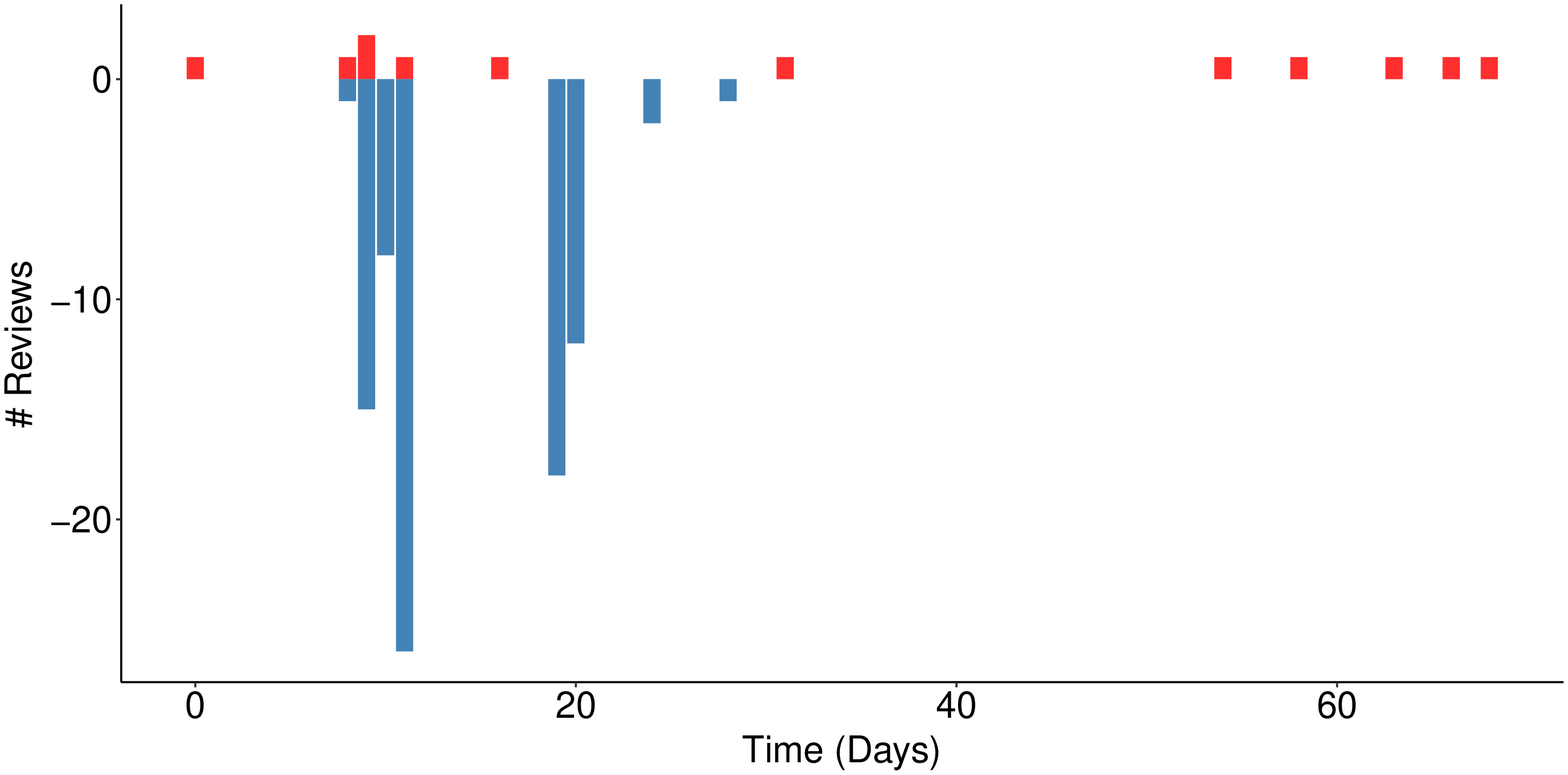}}
\subfigure[]
{\label{fig:timeline_mix_reviews}\includegraphics[width=0.32\textwidth]{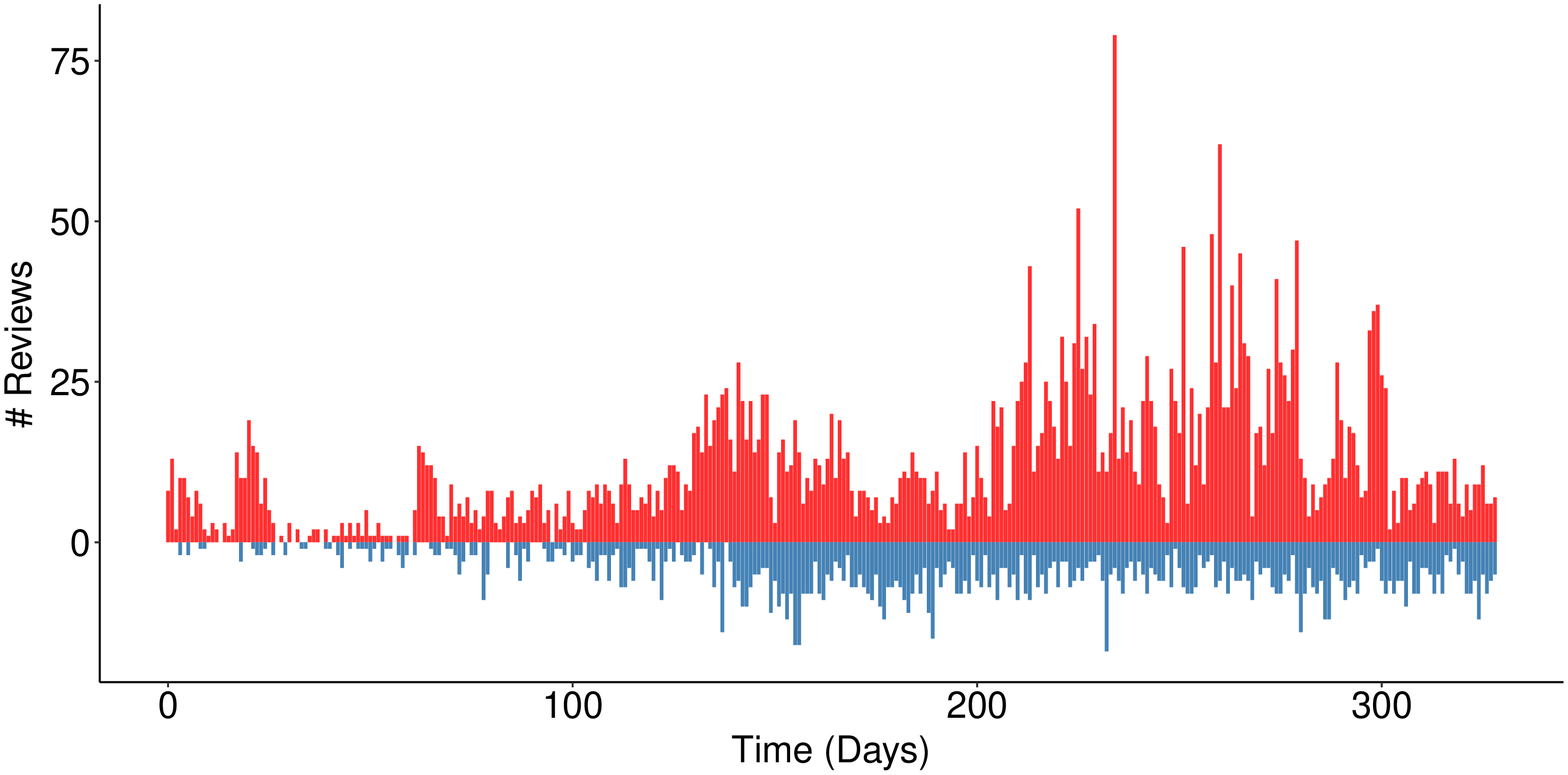}}
\caption{Review timeline of ``fraud'' apps: $x$ axis shows time with a day
granularity, $y$ axis the number of daily positive reviews (red, positive
direction) and negative reviews (blue, negative direction). {\bf Apps can be
targets of both positive and negative search rank fraud campaigns}:
(a) The app ``Daily Yoga- Yoga Fitness Plans'' had days with above 200 positive
review spikes.
(b) ``Real Caller'' received suspicious negative review spikes from
ground truth fraudster-controlled accounts.
(c) ``Crownit - Cashback \& Prizes'' received both positive and negative
reviews from fraudster-controlled accounts.}
\label{fig:app:timeline}
\end{figure*}

This intuition is difficult to verify without access to Google's ranking
function. To compare the dynamics between the top-24 positions and bottom 25,
we computed $M^{k_{t_1},k_{t_2}}$, the amount of overlap between two subsequent
lists for the two cases (see Table~\ref{tab:mvalue}). In addition, we also
computed the overlap between the first and the last lists obtained over the
observation period. The overlap between the first and the last observed lists
is zero in the case of the \textit{New(Free)} and \textit{New(Paid)} lists, due
to the higher in-flow of apps in these lists.

\begin{table}
\centering
\resizebox{0.49\textwidth}{!}{%
\textsf{
\begin{tabular}{|l| l| r|r|r|r|r|r|}
\hline
Pages & List-Type & \#items & $O_{mean}$ & $O_{min}$ & $M_{mean}$ & $M_{sd}$ & $O_{f \cap l}$\tabularnewline
\hline
\hline
\multirow{5}{0.35in}{\textbf{Top 24}} & Gross & 57 & 23.8953 & 18 & 0.9893 & 0.0617 & 12\tabularnewline
\cline{2-8}
 & Free & 44 & 23.9539 & 21 & 0.9970 & 0.0238 & 16\tabularnewline
\cline{2-8}
 & Paid & 74 & 23.8840 & 11 & 0.9932 & 0.0537 & 4\tabularnewline
\cline{2-8}
 & New (Free) & 128 & 23.7974 & 15 & 0.9867 & 0.0742 & 0\tabularnewline
\cline{2-8}
 & New (Paid) & 125 & 23.8226 & 13 & 0.9889 & 0.0641 & 0\tabularnewline
\cline{1-8}
\multirow{5}{0.35in}{\textbf{Last 25}} & Gross & 205 & 25.3765 & 1 & 0.9692 & 0.1299 & 0\tabularnewline
\cline{2-8}
 & Free & 186 & 25.6145 & 10 & 0.9785 & 0.1030 & 4\tabularnewline
\cline{2-8}
 & Paid & 150 & 24.9245 & 12 & 0.9780 & 0.1029 & 4\tabularnewline
\cline{2-8}
 & New (Free) & 449 & 25.1159 & 5 & 0.9571 & 0.1502 & 0\tabularnewline
\cline{2-8}
 & New (Paid) & 485 & 24.9245 & 2 & 0.9780 & 0.1687 & 0\tabularnewline
\cline{1-8}
\hline
\end{tabular}}}
\caption{
\normalfont{
Variability in top-k Lists. $O_{mean}$, $O_{min}$ and
$O_{f \cap l}$ are the mean and min. overlap, and that
between the first and last lists.}
}
\label{tab:mvalue}
\vspace{-15pt}
\end{table}

In all other cases (except \textit{Paid} top-24 and last-25
cases), there is an overlap of at least 50\%: apps continue
to be popular for longer periods.  In each list-type of last-25 cases, the low
overlap values indicate that the list is highly dynamic and variable.  Also,
notice that $M_{mean}$ for the top-24 is higher than that of the last-25
indicating that the top-24 is less dynamic in all cases expect
\textit{New(Free)} and \textit{New(Paid)}.

Figure~\ref{fig:topk_M-topk_occupants} shows the number of apps that occupy a
rank position in 5 different list-types over our observation period. Note that
a \textit{lower rank is preferred}.  For example, the $300^{th}$ rank position
in the \textit{New (Free)} list is occupied by 441 applications. With the
increase in rank, the rate of applications being swapped is increasing for each
category indicating an increased churn -- it is easier for apps to occupy as
well as get displaced on high ranks.  For \textit{Paid}, \textit{Gross}, and
\textit{Free}, the number of apps varies from 34 to 142, 30 to 173, and
43 to 163, respectively, from the $1^{st}$ to the $400^{th}$ rank.

However, in the case of \textit{New (Free)} and \textit{New(Paid)} lists, the
number of apps being swapped for a position is almost linearly increasing with
the increase in rank.  This is because all the applications in these two
categories are new and the competition is higher compared to other list-types.


Figure~\ref{fig:topk_M-topk_lifetime_densityplot} shows the distribution of the
lifetime of applications that occupy a specific rank position. To evaluate the
variation in the distributions we choose the $1^{st}$, $50^{th}$, $100^{th}$,
$200^{th}$, and $400^{th}$ rank positions. For each category, the average
lifetime is longer for higher ranking apps then for lower ranking apps. We can
clearly observe this phenomenon in the case of \textit{New (Free)} and
\textit{New (Paid)}. In both the cases, the lifetime of the apps at the lowest
rank (\textit{i.e.,} $400^{th}$) is the lowest, \textit{i.e.}, $\approx$6 hours
and it starts increasing with the increase in the ranks. We can attribute this
effect to the frequently changing list of new apps and the relatively easier
competitions to be on the top-400 lists.  However, in case of \textit{Free}
apps, the average lifetime of apps on the $1^{st}$ rank is $94.2$ hours and
decreases to $16.7$ hours for the $400^{th}$ rank. For \textit{Paid} and
\textit{Gross} categories, the lifetime changes from $81.6$ to $20.6$ hours and
$65.7$ to $17.9$ hours, respectively, for the rank $1 \rightarrow 400$.  We
attribute these effects to the stability of the apps in these lists.


\section{Research Implications}
\label{sec:discussion}

\noindent
We now discuss the implications of longitudinal monitoring on security and
systems research in Android app markets.

\subsection{Fraud and Malware Detection}
\label{sec:discussion:threat}

App markets play an essential role in the profitability of apps. Apps ranked
higher in the app market become more popular, thus make more money, either
through direct payments for paid apps, or through ads for free apps. This
pressure to succeed leads some app developers to tinker with app market
statistics known to influence the app ranking, e.g., reviews, average rating,
installs~\cite{J13}. Further, malicious developers also attempt to use app
markets as tools to widely distribute their malware apps.  We conjecture that a
longitudinal analysis of apps can reveal both fraudulent and malicious apps. 
In the following we provide supporting evidence.

\noindent
{\bf Search rank fraud}.
We have contacted Freelancer workers specializing in Google Play fraud, and
have obtained the ids of 2,600 Google Play accounts that were used to write
fraudulent reviews for 201 unique apps. We have analyzed these apps and found
that fraudulent app search optimization attempts often produce suspicious
review patterns. A longitudinal analysis of an app's reviews, which we call
{\it timeline}, can reveal such patterns. For instance,
Figure~\ref{fig:timeline_pos_reviews} shows the review timeline of ``Daily
Yoga- Yoga Fitness Plans'', one of the 201 apps targeted by the 15
fraudster-controlled accounts. We observe several suspicious positive reviews
spikes, some at over 200 reviews per day, in contrast with long intervals
of under 50 daily positive reviews.

We have observed that Google Play apps can also be the target of negative
review campaigns, receiving negative reviews from multiple fraudster-controlled
accounts.  Figure~\ref{fig:timeline_neg_reviews} shows the timeline of such an
app, ``Real Caller'', where we observe days with up to 25 negative reviews, but
few positive reviews. While negative reviews are often associated with poor
quality apps, these particular spikes are generated from the
fraudster-controlled accounts mentioned above. We conjecture that negative
review campaigns are sponsored by competitors. Further, we identified apps that
are the target of both positive and negative reviews.
Figure~\ref{fig:timeline_mix_reviews} shows the timeline of such an app,
``Crownit - Cashback \& Prizes''. While the app has received more positive
reviews with higher spikes, its negative reviews and spikes thereof are also
significant.

App markets can monitor timelines and notify developers and their users when
such suspicious spikes occur.


In addition,
our analysis has shown that several developers upload many unpopular apps (see
$\S\ref{sec:developer:impact}$), while others tend to push frequent updates
($\S$\ref{sec:pop:staleness}). We describe here vulnerabilities related to
such behaviors.

\noindent
\textbf{Scam Apps.}
We have identified several ``productive'' developers, that upload many similar
apps. Among them, we have observed several thousands of premium applications
(priced around \$1.99) that are slight variations of each other and have almost
no observable functionality.  Such apps rely on their names and description to
scam users into paying for them, then fail to deliver. Each such app receives
$\approx$500-1000 downloads, bringing its developer a profit of \$1000-2000.

\begin{figure*}
\centering
\subfigure[]
{\label{fig:timeline_perm2}\includegraphics[width=0.32\textwidth]{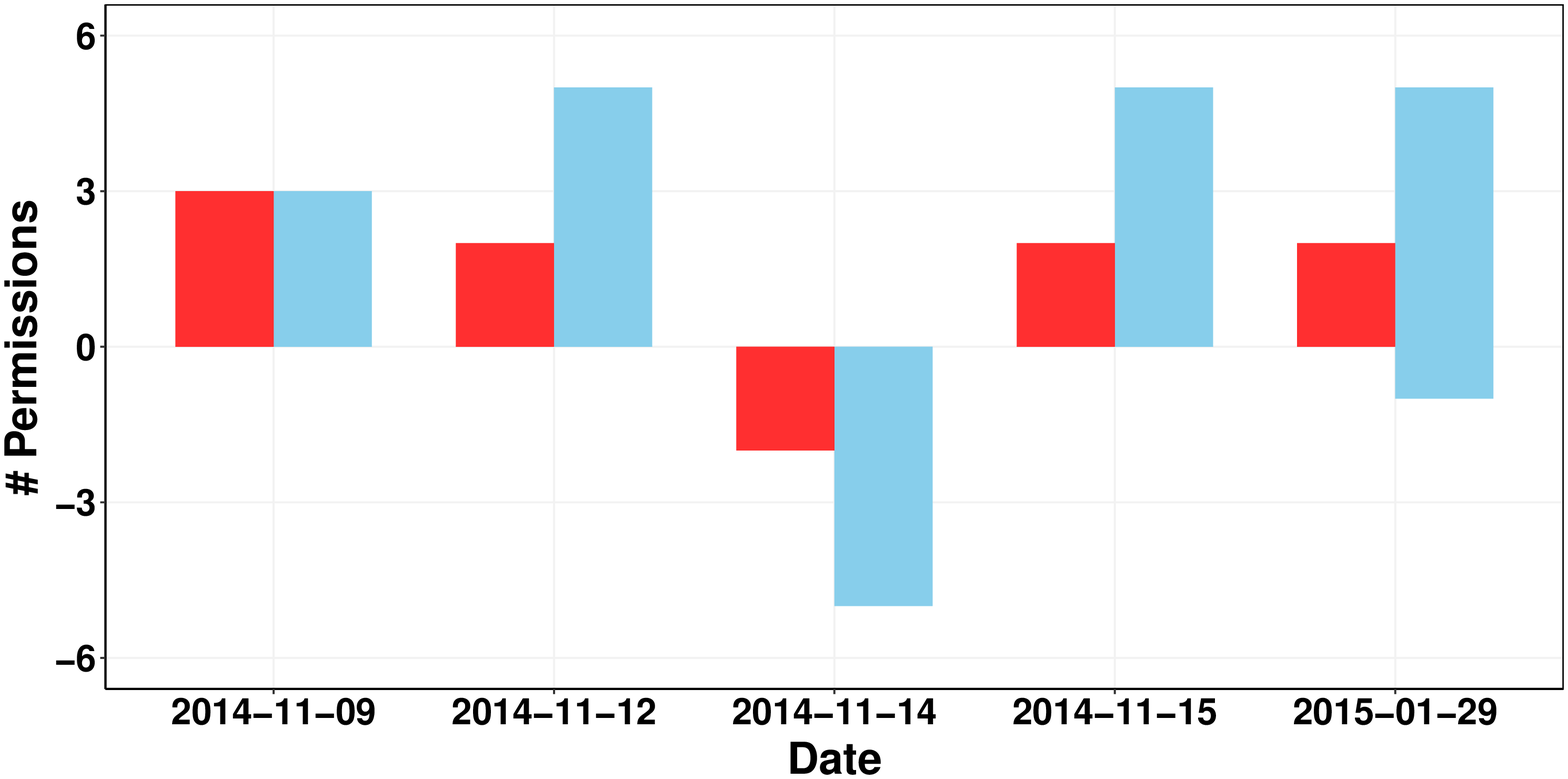}}
\subfigure[]
{\label{fig:timeline_perm3}\includegraphics[width=0.32\textwidth]{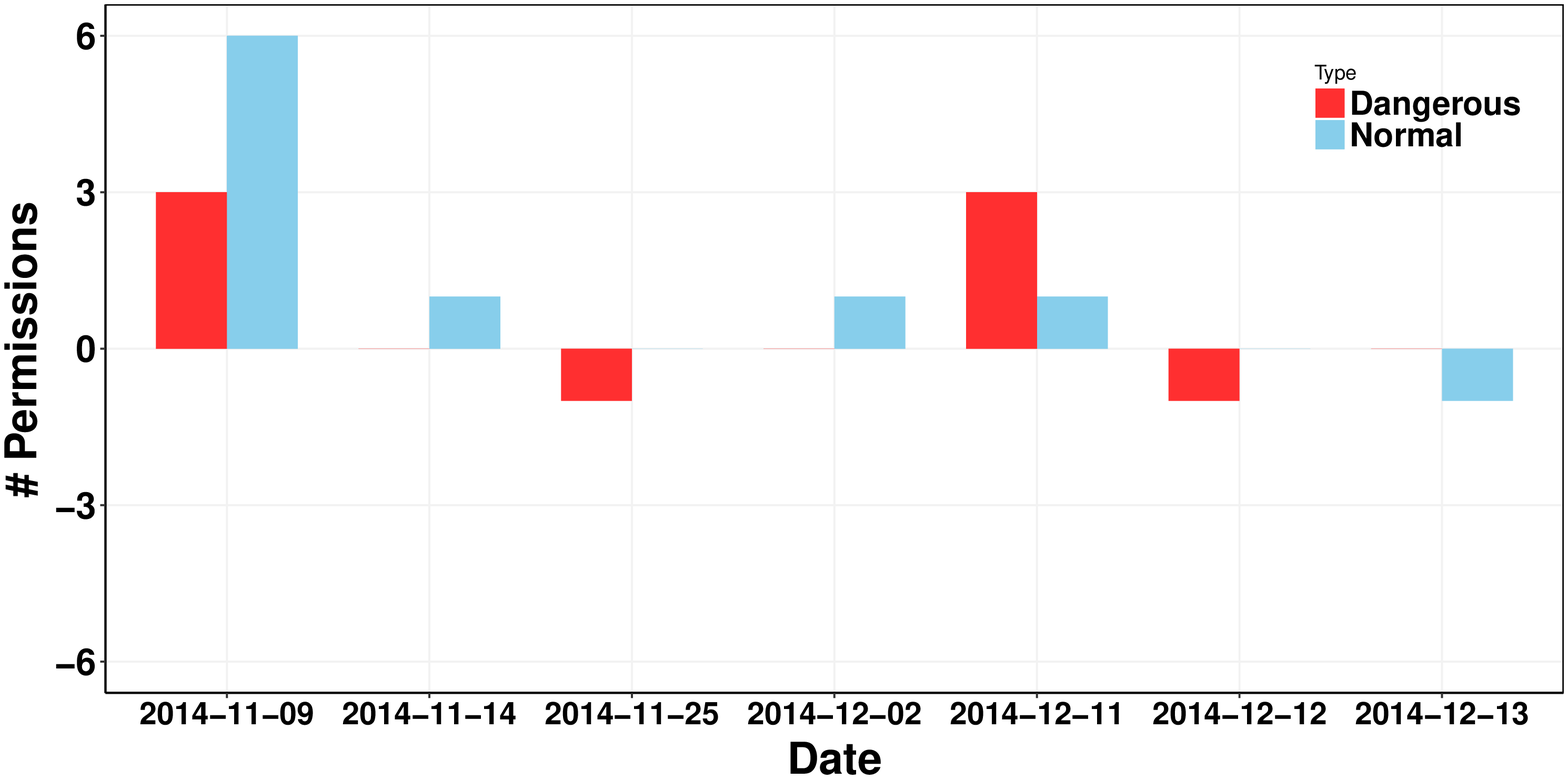}}
\subfigure[]
{\label{fig:timeline_perm1}\includegraphics[width=0.32\textwidth]{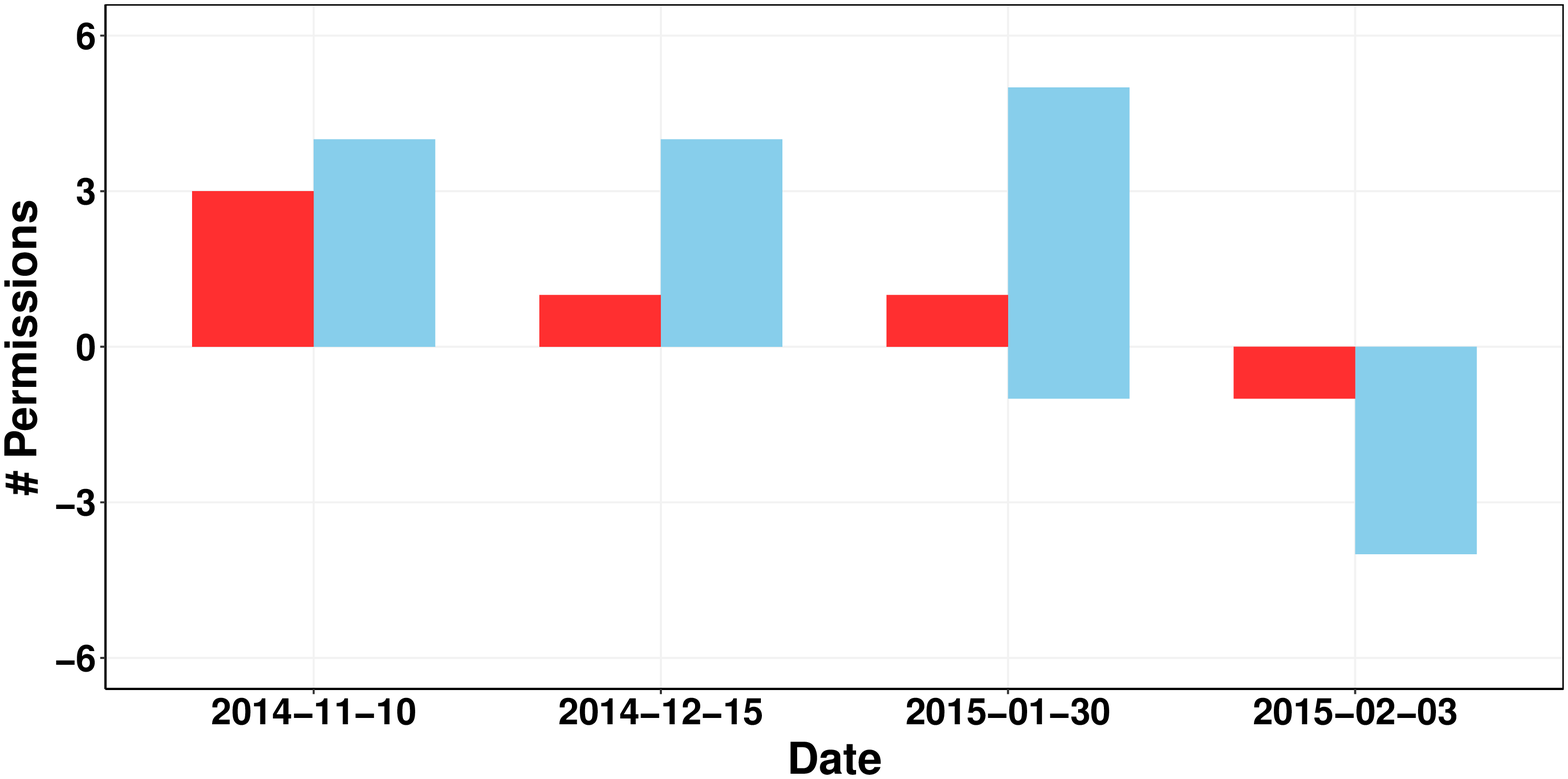}}
\caption{Permission timeline of 3 VirusTotal flagged apps,
(a) ``Hidden Object Blackstone'',
(b) ``Top Race Manager'', and
(c) ``Cash Yourself''.
The $x$ axis shows the date when the permission changes occurred; the $y$ axis
shows the number of permissions that were newly requested (positive direction)
or removed (negative direction). The red bars show dangerous permissions, blue
bars regular permissions. We observe significant permission changes, even
within days.}
\label{fig:app:timeline_permission}
\vspace{-15pt}
\end{figure*}

\noindent
\textbf{Malware}.
While updates enable developers to fix bugs and push new functionality in a
seamless manner, attack vectors can also leverage them. Such attack vectors can
be exploited both by malicious developers and by attackers that infiltrate
developer accounts. We posit that a motivated attacker can develop and upload a
benign app, and once it gains popularity, push malware as an update.  For
instance, Table~\ref{tab:prob-table} shows that as expected, software version
and total permissions are highly correlated. However, we found that in 5\% of
cases where permissions change, the software version does not change.

On the iOS platform, Wang et al.~\cite{WLLCL13} proposed to make the app
remotely exploitable, then introduce malicious control flows by rearranging
already signed code. We propose an Android variant where the attacker ramps up
the permissions required by the app, exploiting the observation that a user is
more likely to accept them, then to uninstall the app.

To provide an intuition behind our conjecture, we introduce the concept of {\it
app permission timeline}, the evolution in time of an app's requests for new
permissions, or decisions to remove permissions. We have used
VirusTotal~\cite{VirusTotal} to test the apks of 7,756 randomly selected apps
from the dataset.14-15. We have selected apps for which VirusTotal raised at
least 3 flags and that have at least 10 reviews.
Figure~\ref{fig:app:timeline_permission} shows the permission timeline of 3 of
these apps, for both dangerous (red bars) and regular permissions (blue bars).

For instance, the ``Hidden Object Blackstone'' app
(Figure~\ref{fig:timeline_perm2}) has a quick succession of permission requests
and releases at only a few days apart. While the app releases 2 dangerous
permissions on November 14, 2014, it requests them again 1 day later, and
requests 2 more a month and a half later. Similarly, the ``Top Race Manager''
app (Figure~\ref{fig:timeline_perm3}) has very frequent permission changes,
daily for the last 3. The ``Cash Yourself'' app
(Figure~\ref{fig:timeline_perm1}) requests 3 dangerous permissions on November
10 2014, followed by 1 dangerous permission in both December and January, then
releases 1 dangerous permission 4 days later.

Permission changes imply significant app changes. Frequent and significant
permission changes, especially the dangerous ones may signal malware, or
unstable apps. Market owners can decide to carefully scan the updates of such
apps for malware, and notify developers that something went wrong with their
updates, indicating potential account infiltration.



\subsection{App Market Ecosystem}

\noindent
\textbf{Analytics-driven Application Development.}
We envision a development model where insights derived from raw market-level
data is integrated into the application development.  Such a model is already
adopted by websites such as Priceline~\cite{priceline} through their ``Name
Your Own Price'' scheme where the interface provides users with hints on
setting an optimal price towards a successful bid. We propose the extension of
development tools like Google's Android Studio~\cite{android-studio} with
market-level analytics, including:

\begin{compactitem}

\item
{\bf Median price}: In $\S$\ref{sec:staleness}, we showed that developers may
be settling down for lower profits. The development tools could provide
developers them with hints on the optimal price for their app based on, e.g.,
the number of features, the price of active apps in the same category etc.

\item
{\bf Application risk}: Provide predictions on the impact of permissions and updates on reviews and download count.

\item
{\bf App insights}: Present actionable insights extracted from user
reviews (e.g., using solutions like NetSieve~\cite{potharaju2013juggling}),
including most requested feature, list of buggy features, features that crash the
app.

\end{compactitem}

\noindent
{\bf Enriching User Experience.}
We believe data-driven insights will be indispensable to enhance the end user
experience:

\begin{compactitem}

\item
{\bf Analytics based app choice}: Visualize app price,
update overhead, required permissions, reviewer
sentiment to enhance the user experience when choosing among apps with similar
claimed functionality. For instance, develop scores for individual features,
and even an overall ``sorting'' score based on user preferences. Scam apps (see
$\S$\ref{sec:discussion:threat}) should appear at the bottom of the score
based sorted app list.

\item
{\bf Analytics based app quarantine}:
We envision a quarantine based approach to defend against ``update'' attacks.
An update installation is postponed until analytics of variation in app
features indicates the update is stable and benign. To avoid a situation where
all users defer installation, we propose a probabilistic quarantine. Each user
can update the app after a personalized random interval after its release.

\end{compactitem}

\section{Limitations}

This paper seeks to shed light on the dynamics of the Google app market and
also provide evidence that a longitudinal monitoring of apps is beneficial for
users, app developers and the market owners. However, our datasets were
collected in 2012 and 2014-2015, and may not reflect the current trends of
Google Play.

In addition, while we believe that the Google Play market, the applications it
hosts and developers we examined represent a large body of other third-party
markets and their environments, we do not intend to generalize our results to
all the smartphone markets. The characteristics and findings obtained in this
study are associated with the Google Play market and its developers. Therefore,
the results should be taken with the market and our data collection methodology
in mind.

The goal of our discussion of permission and review timelines was to provide
early evidence that a longitudinal monitoring and analysis of apps in app
markets can be used to identify suspicious apps.  We leave for future work a
detailed study of permission changes to confirm their statistical significance
in detecting search rank fraud and malware.


\section{Conclusion}
\label{sec:conclusion}

This article studies temporal patterns in Google Play, an influential app
market. We use data we collected from more than 160,000 apps daily over a six
month period, to examine market trends, application characteristics and
developer behavior in real-world market settings. Our work provides insights
into the impact of developer levers (e.g., price, permissions requested, update
frequency) on app popularity. We proposed future directions for integrating
analytics insights into developer and user experiences. We introduced novel
attack vectors on app markets and discussed future detection directions.


\section{Acknowledgments}

This research was supported in part by NSF grants CNS-1527153, CNS-1526494 and
CNS-1450619.

\bibliographystyle{abbrv}
\bibliography{bogdan,references,social.fraud,malware}

\vspace{-25pt}

\begin{IEEEbiographynophoto}
{Rahul Potharaju} is a researcher at Microsoft. He focuses on building
interactive query engines for big data. He earned his CS Ph.D. degree from
Purdue University and CS Master’s degree from Northwestern University. He is a
recipient of the Motorola Engineering Excellence award in 2009, the Purdue
Diamond Award in 2014, and the Microsoft Trustworthy Reliability Computing
Award in 2013. 
\end{IEEEbiographynophoto}

\vspace{-25pt}

\begin{IEEEbiographynophoto}
{Mizanur Rahman} is a Ph.D. candidate at FIU.  He has previously held various
positions in KAZ Software, iAppDragon and Prolog Inc.  His research interests
include fraud detection in social networks and user experience.
\end{IEEEbiographynophoto}

\vspace{-25pt}

\begin{IEEEbiographynophoto}
{Bogdan Carbunar} is an assistant professor in SCIS at FIU. He has held
research positions within Motorola Labs.  His interests include security and
privacy for mobile and social networks.  He holds a Ph.D. in CS from Purdue.
\end{IEEEbiographynophoto}

\end{document}